\documentclass[11pt]{article}
\usepackage{geometry} % see geometry.pdf on how to lay out the page. There's lots.
\usepackage[english]{babel}
\usepackage[english]{layout}

\usepackage{mathpazo} % math & rm
\usepackage[scaled]{helvet} % ss
\usepackage{courier} % tt
\normalfont
\usepackage[T1]{fontenc}

\usepackage{array} % for better arrays (eg matrices) in maths
\usepackage{paralist} % very flexible & customisable lists (eg. enumerate/itemize, etc.)
\usepackage{verbatim} % adds environment for commenting out blocks of text & for better verbatim
\usepackage[utf8]{inputenc} % Any characters can be typed directly from the keyboard, eg éçñ
\usepackage{textcomp} % provide lots of new symbols
\usepackage{graphicx}  % Add graphics capabilities
\usepackage{epstopdf} % to include .eps graphics files with pdfLaTeX
\usepackage{flafter}  % Don't place floats before their definition

\usepackage{amsfonts,latexsym,amsthm,amssymb}
\usepackage{amsmath,amssymb,bbm,amsfonts}  % Better maths support & more symbols
\usepackage{bm}  % Define \bm{} to use bold math fonts

\usepackage{memhfixc}  % remove conflict between the memoir class & hyperref
\usepackage{fancyhdr} % This should be set AFTER setting up the page geometry
\usepackage{sectsty}
\allsectionsfont{\sffamily\mdseries\upshape} % (See the fntguide.pdf for font help)
\numberwithin{equation}{section}

\usepackage{color}

\begin{document}

\title{\LARGE \textbf{Algebraic Bethe ansatz for the XXX chain with triangular boundaries and Gaudin model}}

\author{\textsf{N. ~Cirilo ~Ant\'onio,}
\thanks{E-mail address: nantonio@math.ist.utl.pt}
\textsf{ ~ N. ~Manojlovi\'c}
\thanks{E-mail address: nmanoj@ualg.pt}
\textsf{ and I. ~Salom}
\thanks{E-mail address: 
isalom@ipb.ac.rs} \\
\\
\textit{$^{\ast}$Centro de An\'alise Funcional e Aplica\c{c}\~oes}\\
\textit{Instituto Superior T\'ecnico, Universidade de Lisboa} \\
\textit{Av. Rovisco Pais, 1049-001 Lisboa, Portugal} \\
\\
%\textit{$^{\dag}$Department of Mathematics, Massachusetts Institute of Technology} \\
%\textit{77 Massachusetts Avenue, Cambridge, MA 02139-4307, USA.} \\
%\\
\textit{$^{\dag}$Grupo de F\'{\i}sica Matem\'atica da Universidade de Lisboa} \\
\textit{Av. Prof. Gama Pinto 2, PT-1649-003 Lisboa, Portugal} \\
\\
\textit{$^{\dag}$Departamento de Matem\'atica, F. C. T.,
Universidade do Algarve} \\
\textit{Campus de Gambelas, PT-8005-139 Faro, Portugal}\\
\\
\textit{$^{\ddag}$Institute of Physics, University of Belgrade}\\
\textit{P.O. Box 57, 11080 Belgrade, Serbia}\\
\\
}
\date{}

%\keywords{Lattice systems; Exactly solvable models; Gaudin model}

\maketitle
\thispagestyle{empty}
%\vspace{5mm}
\begin{abstract}
We implement fully the algebraic Bethe ansatz for the XXX Heisenberg spin chain in the case when both boundary matrices can be brought to the upper-triangular form. We define the Bethe vectors
which yield the strikingly simple expression for the off shell action of the transfer matrix, deriving the spectrum and the corresponding Bethe equations. We explore further these results by obtaining the off shell action of the generating function of the Gaudin Hamiltonians on the Bethe vectors through the so-called quasi-classical limit.

\end{abstract}

\clearpage
\newpage

\section{Introduction}
The quantum inverse scattering method (QISM) is an approach to construct and solve quantum integrable systems \cite{TakhtajanFaddeev79, KulishSklyanin82, Faddeev98}. In the framework of the QISM the algebraic Bethe ansatz (ABA) is a powerful algebraic tool, which yields the spectrum and corresponding eigenstates for which highest weight type representations are relevant, like for example quantum spin systems, Gaudin models, etc. In particular, the Heisenberg spin chain \cite{Heisenberg28}, with periodic boundary conditions, has been studied by the algebraic Bethe ansatz \cite{TakhtajanFaddeev79,Faddeev98}, including the question of completeness and simplicity of the spectrum \cite{Tarasov09}.

A way to introduce non-periodic boundary conditions compatible with the integrability of the quantum systems solvable by the  quantum inverse scattering method was developed in \cite{Sklyanin88}. The boundary conditions at the left and right sites of the system are expressed in the left and right reflection matrices. The compatibility condition between the bulk and the boundary of the system takes the form of the so-called reflection equation. The compatibility at the right site of the model is expressed by the dual reflection equation. The matrix form of the exchange relations between the entries of the Sklyanin monodromy matrix are analogous to the reflection equation. Together with the dual reflection equation they yield the commutativity of the open transfer matrix \cite{Sklyanin88, FreidelMaillet91, FreidelMaillet91a}.

There is a renewed interest in applying the algebraic Bethe ansatz to the open XXX chain with non-periodic boundary conditions compatible with the integrability of the systems \cite{Martins05, Eric13, Belliard13, Lima13}. Other approaches include the ABA based on the functional relation between the eigenvalues of the transfer matrix and the quantum determinant and the associated T-Q relation \cite{China13}, functional relations  for the eigenvalues of the transfer matrix based on fusion hierarchy \cite{Nepomechie04} and the Vertex-IRF correspondence \cite{China03}. For a review of the coordinate Bethe ansatz for non-diagonal boundaries see \cite{Eric2013RMP}. However, we will focus on the case when system admits the so-called pseudo-vacuum, or the reference state \cite{Sklyanin88, Martins05, Eric13, Belliard13, Lima13}. In his seminal work on boundary conditions in quantum integrable models Sklyanin has studied the XXZ spin chain with diagonal boundaries \cite{Sklyanin88}. The next relevant step was the study of the $s\ell(n)$ spin chain in the case when reflection matrices can be brought into the diagonal form by a suitable similarity transformation which leaves the R-matrix invariant and it is independent of the spectral parameter \cite{French04, Martins05a}. These results were then generalized to the case of the spin-s XXX chain when there exists a basis in which one reflection matrix is triangular and the other one is diagonal \cite{Martins05}. Recent studies are focused on the XXX chain when both K-matrices can be simultaneously brought to a triangular form by a single similarity matrix which is independent of the spectral parameter \cite{ Eric13} and similarly for the XXZ chain \cite{Lima13}. \break Although the on shell Bethe ansatz is realized, the proposed Bethe vectors are not suitable for the off shell ABA. The case when the 
reflection matrix $K^- (\lambda)$ is diagonal and $K^+ (\lambda)$ is a two-by-two matrix with non-zero entries was studied in \cite{Belliard13}.

This work is centred on the implementation of the algebraic Bethe ansatz which yields the off shell action of the transfer matrix the XXX Heisenberg spin chain when the corresponding K-matrices are triangularizable. The Bethe vectors $\Psi _M ( \mu_1 ,  \mu_2 ,  \dots ,  \mu_M )$ we define here are such that they make the off shell action of the transfer matrix \break strikingly simple since it almost coincides with the corresponding action in the case when the two boundary matrices are diagonal.  
The Bethe vectors $\Psi _M ( \mu_1 ,  \mu_2 ,  \dots ,  \mu_M )$, for an arbitrary positive integer $M$, 
are defined explicitly as some polynomial functions of the creation operators. As expected, the off shell action yields the spectrum of the transfer matrix and the corresponding Bethe equations. To explore further these results we use the so-called quasi-classical limit and obtain the off shell action of the generating function of the Gaudin Hamiltonians, with boundary terms, on the corresponding Bethe vectors. 

A model of interacting spins in a chain was first considered by Gaudin \cite{Gaudin76,Gaudin83}. In his approach, these models were introduced as a quasi-classical limit of the integrable quantum chains. The Gaudin models were extended to any simple Lie algebra, with arbitrary irreducible representation at each site of the chain \cite{Gaudin83}. Sklyanin studied the rational $s\ell(2)$ model in the framework of the quantum inverse scattering method using the $s\ell(2)$ invariant classical r-matrix \cite{Sklyanin89}. A generalization of these results to all cases when skew-symmetric r-matrix satisfies the classical Yang-Baxter equation \cite{BelavinDrinfeld} was relatively straightforward \cite{SklyaninTakebe,Semenov97}. Therefore, considerable attention has been devoted to Gaudin models corresponding to the the classical r-matrices of simple Lie algebras \cite{Jurco89, Jurco90, WagnerMacfarlane00} and Lie superalgebras \cite{BrzezinskiMacfarlane94, KulishManojlovic01, KulishManojlovic03, LimaUtiel01, KurakLima04}. 

Hikami showed how the quasi-classical expansion of the transfer matrix, calculated at the special values of the spectral parameter, yields the Gaudin Hamiltonians in the case of non-periodic boundary conditions \cite{Hikami95}. Then the ABA was applied to open Gaudin model in the context of the the Vertex-IRF correspondence \cite{YangZhang12, YangZhangSasakic04, YangZhangSasakic05}. Also,  results were obtained for the open Gaudin models based on Lie superalgebras \cite{Lima09}. An approach to study the open Gaudin models based on the classical reflection equation \cite{Sklyanin86}
and the non-unitary r-matrices was developed recently, see \cite{Skrypnyk09, Skrypnyk13} and the references therein. For a recent review of the open Gaudin model see \cite{CAMN}.

In \cite{CAMRS} we have derived the generating function of the Gaudin Hamiltonians with boundary terms following Sklyanin's approach in the periodic case \cite{Sklyanin89}. Our derivation is based on the quasi-classical expansion of the linear combination of the transfer matrix of the XXX chain and the central element, the so-called Sklyanin determinant. Here we use this result with the objective to derive the off shell action of the generating function of the Gaudin Hamiltonians. As we will show below, the quasi-classical expansion of the Bethe vectors we have defined for he XXX Heisenberg spin chain yields the Bethe vectors of the corresponding Gaudin model. The significance of these Bethe vectors is in the striking simplicity of the formulae of the off shell action of the generating function of the Gaudin Hamiltonians.

This paper is organized as follows. In Section 2 we review the $SL(2)$-invariant Yang R-matrix and provide fundamental tools for the study of the inhomogeneous XXX Heisenberg spin chain. The general solutions of the reflection equation and the dual reflection equation are given in Section 3 as well as the triangularization of these K-matrices, when the corresponding parameters obey an extra identity. In Section 4 we expose the Sklyanin approach to the inhomogeneous XXX Heisenberg spin chain with non-periodic boundary conditions. The implementation of the ABA, as one of the main results of the paper, is presented in Section 5, including the definition of the Bethe vectors and the formulae of the off shell action of the transfer matrix. Corresponding Gaudin model and the respective implementation of the ABA are given in Section 6. Our conclusions are presented in the
Section 7. Finally, in Appendix A are given some basic definitions for the convenience of the reader and in Appendix B are given commutation relations relations relevant for the  implementation of the ABA in Section 5.

\section{Inhomogeneous Heisenberg spin chain }
The XXX Heisenberg  spin chain is related to the Yangian $\mathcal{Y}(s\ell(2))$ (see \cite{ChariPressley}) and the $SL(2)$-invariant Yang R-matrix \cite{Yang67}
\begin{equation}
\label{YangR}
R (\lambda) = \lambda \mathbbm{1}  + \eta \mathcal{P } = \left(\begin{array}{cccc}
\lambda + \eta & 0 & 0 & 0 \\
0 & \lambda & \eta & 0 \\
0 & \eta & \lambda  & 0 \\
0 & 0 & 0 & \lambda + \eta \end{array}\right),
\end{equation}
where $\lambda$ is a spectral parameter, $\eta$ is a quasi-classical parameter. We use $\mathbbm{1}$ for the identity operator and $\mathcal{P }$ for the permutation in $\mathbb{C} ^2 \otimes \mathbb{C} ^2$.

The Yang R-matrix satisfies the Yang-Baxter equation \cite {Yang67, Baxter82} in the space $\mathbb{C}^2 \otimes \mathbb{C}^2 \otimes \mathbb{C}^2$
\begin{equation}
\label{YBE}
R_{12} ( \lambda - \mu) R_{13} ( \lambda) R_{23} (\mu) = R_{23} (\mu ) R_{13} (\lambda ) R_{12} ( \lambda - \mu),
\end{equation}
we suppress the dependence on the quasi-classical parameter $\eta$ and use the standard notation of the QISM to denote spaces $V_j, j=1, 2, 3$ on which corresponding $R$-matrices $R_{ij}, ij = 12, 13, 23$ act non-trivially \cite{TakhtajanFaddeev79, KulishSklyanin82, Faddeev98}. In the present case $V_1 = V_2 = V_3 =\mathbb{C}^2$.

The Yang R-matrix also satisfies other relevant properties such as

\begin{tabbing}{|l}
xxxxxxxxxxxxxxxxxxxxxxxxxx   \= xxxxxxxxxxxxxxxxxxxxxxxxxx     \kill
unitarity    \>   $R_{12} ( \lambda ) R_{21} ( -\lambda ) = (\eta ^2 - \lambda ^2 )  \mathbbm{1}$;    \\
parity invariance    \> $R_{21} ( \lambda ) = R_{12} ( \lambda ) $;  \\
temporal invariance  \> $R_{12}^t ( \lambda) = R_{12} ( \lambda)$; \\
crossing symmetry  \>  $R ( \lambda) =   \mathcal{J} _1 R ^{t_2}( -\lambda - \eta ) {\mathcal{J}}^{-1} _1,$
\end{tabbing}
where $t_2$ denotes the transpose in the second space and the entries of the two-by-two matrix $\mathcal{J}$ are $\mathcal{J}_{ab}=  (-1)^{a-1}\delta_{a,3-b}$.

Here we study the inhomogeneous XXX spin chain with $N$ sites, characterised by the local space 
$V_ m = \mathbb{C}^{2s+1}$ and inhomogeneous parameter $\alpha _m$. The Hilbert space of the system is
\begin{equation}
\label{H-space}
\mathcal{H} = \underset {m=1}{\overset {N}{\otimes}}  V_m = (\mathbb{C}^{2s+1} ) ^{\otimes N}.
\end{equation}
Following \cite{Sklyanin89} we introduce the Lax operator
\begin{equation}
\label{L-XXX}
\mathbb{L}_{0m}(\lambda) 
%=  \mathbbm{1} + \frac{\eta}{\lambda} \left( \mathcal{P}_{0m} -  \frac{1}{2} \mathbbm{1} \right) 
= \mathbbm{1} + \frac{\eta}{\lambda} \left( \vec{\sigma}_{0} \cdot \vec{S}_{m} \right) 
= \frac{1}{\lambda}  \left(\begin{array}{cc}  
\lambda + \eta S_m^{3} & \eta  S_m^{-} \\ \eta S_m^{+} & \lambda - \eta S_m^{3}
\end{array}\right).
\end{equation}
Notice that $\mathbb{L}(\lambda)$ is a two-by-two matrix in the auxiliary space $V_0 = \mathbb{C}^2$.
It obeys
\begin{equation} 
\label{unit:Lax}
\mathbb{L}_{0m}(\lambda) \mathbb{L}_{0m}(\eta-\lambda) 
= \Big(1+ \eta^2 \, \frac{s_m (s_m+1)}{\lambda(\eta-\lambda)}\Big) \mathbbm{1}_0\,,
\end{equation}
where $s_m$ is the value of spin in the space $V_m$.

When the quantum space is also a spin $\frac12$ representation, the Lax operator becomes the $R$-matrix,
$\mathbb{L}_{0m}(\lambda) = \frac{1}{\lambda} R_{0m}\left(\lambda - {\eta}/{2} \right) $.

Due to the commutation relations \eqref{crspin1}, it is straightforward to check that the Lax operator satisfies the RLL-relations
\begin{equation}
\label{RLL}
R_{00'} ( \lambda - \mu) \mathbb{L}_{0m}( \lambda - \alpha _m ) \mathbb{L}_{0'm}( \mu - \alpha _m ) =  \mathbb{L}_{0'm}( \mu  - \alpha _m ) \mathbb{L}_{0m}( \lambda  - \alpha _m )R_{00'} ( \lambda - \mu).
\end{equation}
The so-called monodromy matrix
\begin{equation}
\label{monodromy-T}
T(\lambda ) = \mathbb{L}_{0N} ( \lambda - \alpha _N) \cdots \mathbb{L}_{01} ( \lambda - \alpha _1)
\end{equation}
is used to describe the system. For simplicity we have omitted the dependence on the quasi-classical parameter $\eta$ and the inhomogeneous parameters $\{ \alpha _j , j = 1 , \ldots , N \}$. Notice that $T(\lambda)$ is a two-by-two matrix acting in the auxiliary space $V_0 = \mathbb{C}^2$, whose entries are operators acting in $\mathcal{H}$
\begin{equation}
\label{T-mat}
T(\lambda ) = \left(\begin{array}{cc} 
A(\lambda ) &  B(\lambda ) \\ 
C(\lambda ) &  D(\lambda ) \end{array}\right) .
\end{equation}
From RLL-relations \eqref{RLL} it follows that the monodromy matrix satisfies the RTT-relations
\begin{equation}
\label{RTT}
R_{00'} ( \lambda - \mu) {T}_{0} (\lambda ) {T}_ {0'}(\mu ) =  {T} _ {0'}(\mu ){T}_ {0}(\lambda ) R_{00'} ( \lambda - \mu).
\end{equation}
The RTT-relations define the commutation relations for the entries of the monodromy matrix.

In every $V_ m = \mathbb{C}^{2s+1}$ there exists a vector $\omega_m \in V_ m$ such that
\begin{equation}
\label{S-on-om}
S^3_m \omega _m = s_m \omega _m  \quad \text{and}  \quad S^+_m \omega _m = 0 .
\end{equation}
We define a vector $\Omega _+$ to be
\begin{equation}
\label{Omega+}
\Omega _+ = \omega _1 \otimes \cdots \otimes \omega _N \in \mathcal{H}.
\end{equation} 
From the definitions above it is straightforward to obtain the action of the entries of the monodromy matrix \eqref{T-mat} on the vector $\Omega_+$ 
\begin{eqnarray}
\label{AonOm}
A(\lambda) \Omega_+  &=&  a (\lambda) \Omega_+ , \quad \text{with}  \quad  a(\lambda) = \prod _{m=1}^N \frac{\lambda - \alpha _m + \eta s_m}{\lambda - \alpha _m} , \\
\label{DonOm}
D(\lambda) \Omega_+  &=&  d (\lambda) \Omega_+ , \quad \text{with}  \quad  d(\lambda) = \prod _{m=1}^N \frac{\lambda - \alpha _m - \eta s_m}{\lambda - \alpha _m} , \\
\label{ConOm}
C(\lambda) \Omega_+  &=& 0.                                              
\end{eqnarray}

To construct integrable spin chains with non-periodic boundary condition, we will follow Sklyanin's approach \cite{Sklyanin88}. Accordingly, before defining the essential operators and corresponding algebraic structure, in the next section we will introduce the relevant boundary K-matrices. 

\section{Reflection equation}
A way to introduce non-periodic boundary conditions which are compatible with the integrability of the bulk model, was developed in \cite{Sklyanin88}. Boundary conditions on the left and right sites of the system are encoded in the left and right reflection matrices $K^-$ and $K^+$. The compatibility condition between the bulk and the boundary of the system takes the form of the so-called reflection equation. It is written in the following form for the left reflection matrix acting on the space $\mathbb{C}^2$ at the first site $K^-(\lambda) \in \mathrm{End} (\mathbb{C}^2)$
\begin{equation}
\label{RE}
R_{12}(\lambda - \mu) K^-_1(\lambda) R_{21}(\lambda + \mu) K^-_2(\mu)=
K^-_2(\mu) R_{12}(\lambda + \mu) K^-_1(\lambda) R_{21}(\lambda - \mu) .
\end{equation}

Due to the properties of the Yang R-matrix the dual reflection equation can be presented in the following form
\begin{equation}
\label{dRE}
R_{12}( \mu-\lambda )K_1^{+}(\lambda) R_{21}(-\lambda - \mu - 2\eta)  K_2^{+}(\mu)=
K_2^{+}(\mu) R_{12}(-\lambda -\mu-2\eta) K_1^{+}(\lambda) R_{21}(\mu-\lambda) .
\end{equation}
One can then verify that the mapping
\begin{equation}
\label{bijectionKpl}
K^+(\lambda)= K^{-}(- \lambda -\eta)
\end{equation}
is a bijection between solutions of the reflection equation and the dual reflection equation. After substitution of \eqref{bijectionKpl} into the dual reflection equation \eqref{dRE} one gets the reflection equation \eqref{RE} with shifted arguments.

The general, spectral parameter dependent, solutions of the reflection equation \eqref{RE} and the dual reflection equation \eqref{dRE} can be written as follows \cite{VegaGonzalez, KMN3}
\begin{align}
\label{K-min}
\widetilde{K} ^{-}(\lambda) &=  \left(\begin{array}{cc}
\xi ^{-} - \lambda & \widetilde{\psi} ^{-} \lambda \\ \widetilde{\phi} ^{-} \lambda & \xi ^{-} + \lambda \end{array}\right) , \\[2ex]
\label{K-pl}
\widetilde{K} ^{+}(\lambda) &= %K^{-}(- \lambda -\eta) = 
\left(\begin{array}{cc}
\xi ^{+} + \lambda + \eta & - \widetilde{\psi} ^{+} (\lambda + \eta) \\ 
- \widetilde{\phi} ^{+} (\lambda + \eta) & \xi ^{+} - \lambda - \eta 
\end{array}\right) .
\end{align}

We notice that the matrix $K^{-}(\lambda)$ \eqref{K-min} has at most two distinct eigenvalues
\begin{equation}
\label{K-min-e-vals}
\epsilon _{\pm} = \xi ^- \pm \lambda \nu ^-, \qquad \nu ^- = \sqrt{1 + \widetilde{\phi} ^- \widetilde{\psi} ^-} ,
\end{equation}
when $\nu ^- \neq 0$. Then, for $\widetilde{\psi} ^- \neq 0$, there exists a matrix
\begin{equation}
\label{U-mat}
U = \left(\begin{array}{cc} \widetilde{\psi }^- & \widetilde{\psi} ^- \\ 1 - \nu ^- & 1 + \nu ^- \end{array}\right)
\end{equation}
such that
\begin{equation}
\label{K-diag}
U ^{-1} \widetilde{K} ^{-}(\lambda) U = \left(\begin{array}{cc} \xi ^- - \lambda \nu ^- & 0 \\ 0 & \xi ^- + \lambda \nu ^-\end{array}\right) .
\end{equation}
A similar diagonalization exists when $\widetilde{\phi} ^- \neq 0$. However, for $\nu ^- = 0$, i.e. $\widetilde{\phi} ^- \widetilde{\psi} ^- = -1$, the matrix $K^{-}(\lambda)$ cannot be diagonalized and  
\begin{equation}
\label{K-trian}
U ^{-1} K^{-}(\lambda) U = \left(
\begin{array}{cc} 
\xi ^- & \lambda \widetilde{\phi} ^- \\ 
0 & \xi ^- 
\end{array}
\right) ,
\end{equation}
where
\begin{equation}
\label{U-matr}
U = \left(\begin{array}{cc} \widetilde{\psi} ^- & 0 \\ 1 &  - \widetilde{\phi} ^-  \end{array}\right) .
\end{equation}

Following \cite{Eric13} we notice the condition 
\begin{equation}
\label{cond}
\left( \widetilde{\phi} ^- \widetilde{\psi} ^+ - \widetilde{\phi} ^+ \widetilde{\psi} ^- \right) ^2 = 4 \left( \widetilde{\phi} ^- - \widetilde{\phi} ^+ \right) \left( \widetilde{\psi} ^- - \widetilde{\psi} ^+ \right) 
\end{equation}
has to be imposed on the parameters of $K ^{\mp}$ so that the matrices \eqref{K-min} and \eqref{K-pl} are upper triangularizable by a single similarity matrix $M$.  When the square root with the negative sign is taken 
on the right-hand-side of \eqref{cond} then one possible choice for $M$ is given by
\begin{equation}
\label{M-mat}
M = \left(\begin{array}{cc}  - 1 - \nu ^- & \widetilde{\phi} ^- \\ \widetilde{\phi} ^-  &  - 1 - \nu ^-  \end{array}\right) .
\end{equation}
Evidently this matrix does not depend on the spectral parameter $\lambda$ and it is such that
\begin{align}
\label{K-minus}
K^{-}(\lambda) &= M^{-1}  \widetilde{K} ^{-}(\lambda) M  =  \left(
\begin{array}{cc}
\xi ^{-} - \lambda \nu ^- & \lambda \psi ^- \\
0 & \xi ^{-} + \lambda \nu ^-
\end{array}
\right) , \\
\label{K-plus}
K ^{+}(\lambda) &= M^{-1} \widetilde{K} ^{+}(\lambda) M = \left(
 \begin{array}{cc}
 \xi ^{+} + ( \lambda + \eta ) \nu ^+ &   - \psi ^{+} (\lambda + \eta) \\
 0 & \xi ^{+} - ( \lambda + \eta ) \nu ^+
 \end{array}
 \right),
\end{align} 
with $\psi ^- = \widetilde{\phi} ^- + \widetilde{\psi} ^-$, $\nu ^+ = \sqrt{1 + \widetilde{\phi} ^+ \widetilde{\psi} ^+}$ and $\psi ^+ = \widetilde{\phi} ^+ + \widetilde{\psi} ^+$. An analogous choice for $M$ exists for the other sign of the square root in \eqref{cond}.

\section{Inhomogeneous Heisenberg spin chain with boundary terms}
In order to develop the formalism necessary to describe an integrable spin chain with non-periodic boundary condition, we use the Sklyanin approach \cite{Sklyanin88}. The main tool in this framework is the corresponding monodromy matrix  
\begin{equation}
\label{calT}
\mathcal{T}_0(\lambda)= T_0(\lambda) K _0^{-}(\lambda) \widetilde T_0(\lambda),
\end{equation}
it consists of the matrix $T(\lambda)$ \eqref{monodromy-T},  a reflection matrix $K ^{-}(\lambda)$ \eqref{K-minus} and the matrix
\begin{equation}
\label{T-til}
\begin{split}
\widetilde {T}_0(\lambda)= 
\left(\begin{array}{cc} 
\widetilde{A}(\lambda ) &  \widetilde{B}(\lambda ) \\ 
\widetilde{C}(\lambda ) &  \widetilde{D}(\lambda ) 
\end{array}\right) = \mathbb{L}_{01}(\lambda + \alpha _1 + \eta) \cdots \mathbb{L}_{0N} (\lambda + \alpha _N + \eta) .
\end{split}
\end{equation}
It is important to notice that the identity \eqref{unit:Lax}
can be rewritten in the form
\begin{equation}
\label{unit:LaxII}
\mathbb{L}_{0m}(\lambda - \alpha _m) \mathbb{L}_{0m}(-\lambda + \alpha _m + \eta) 
= \Big(1+\frac{\eta^2\,s_m (s_m+1)}{(\lambda - \alpha _m) (-\lambda + \alpha _m + \eta) } \Big) \mathbbm{1} _0 \, .
\end{equation}
It follows from the equation above and the RLL-relations \eqref{RLL} that the RTT-relations \eqref{RTT} can be recast as follows
\begin{align}
\label{tTRT}
\widetilde{T} _ {0'}(\mu ) R_{00'} ( \lambda + \mu) T _{0} (\lambda )  &= T _ {0}(\lambda ) R_{00'} ( \lambda + \mu) \widetilde{T} _ {0'}(\mu ) , \\
\label{tTtTR}
\widetilde{T} _{0} (\lambda ) \widetilde{T} _ {0'}(\mu ) R_{00'} (\mu - \lambda)   &= R_{00'} (\mu - \lambda) \widetilde{T} _ {0'}(\mu ) \widetilde{T} _ {0}(\lambda ) .
\end{align}
Using the RTT-relations \eqref{RTT}, \eqref{tTRT}, \eqref{tTtTR} and the reflection equation \eqref{RE} it is straightforward to show that the exchange relations of the monodromy matrix $\mathcal{T}(\lambda)$  in $V_0\otimes V_{0'}$ are
\begin{equation}
\label{exchangeRE}
R _{00'}(\lambda - \mu) \mathcal{T}_{0} (\lambda) R _{0'0} (\lambda + \mu) \mathcal{T} _{0'} (\mu) = 
\mathcal{T}_{0'}(\mu) R _{00^{\prime}} (\lambda + \mu) \mathcal{T}_{0} (\lambda) 
R _{0'0} (\lambda - \mu) ,
\end{equation}
using the notation of \cite{Sklyanin88}. From the equation above we can read off the commutation relations of the entries of the monodromy matrix
\begin{equation}
\label{calT-mat}
\mathcal{T}(\lambda) = \left(\begin{array}{cc}
\mathcal{A} (\lambda) & \mathcal{B} (\lambda)  \\ 
\mathcal{C} (\lambda) & \mathcal{D} (\lambda)  \end{array} \right) .
\end{equation}
Following Sklyanin \cite{Sklyanin88} (see also \cite{Eric13}) we introduce the operator
\begin{equation}
\label{D-hat}
\widehat{\mathcal{D}} (\lambda) = \mathcal{D} (\lambda) - \frac{\eta}{2\lambda + \eta} \mathcal{A} (\lambda) .
\end{equation}
The relevant commutation relations are given in the appendix B.

The exchange relations \eqref{exchangeRE} admit a central element,  the so-called Sklyanin determinant,
\begin{equation}
\label{Delta-T-cal}
\Delta \left[\mathcal{T}(\lambda)\right] = \mathrm{tr}_{00'} P^{-}_{00'} \mathcal{T}_{0}(\lambda-\eta/2) R_{00'} (2\lambda) \mathcal{T}_{0'}(\lambda+\eta/2). 
\end{equation}
The element $\Delta \left[\mathcal{T}(\lambda)\right]$ can be expressed in form
\begin{equation}
\label{Del-calT}
\Delta \left[\mathcal{T}(\lambda)\right] = 2 \lambda \widehat{\mathcal{D}}  (\lambda - \eta/2) \mathcal{A} (\lambda + \eta/2 )  - (2 \lambda + \eta ) \mathcal{B} (\lambda - \eta/2)  \mathcal{C} (\lambda + \eta/2 ).
\end{equation}

The open chain transfer matrix is given by the trace of the monodromy $\mathcal{T}(\lambda)$ over the auxiliary space $V_0$ with an extra reflection matrix $K^+(\lambda)$ \cite{Sklyanin88},
\begin{equation}
\label{open-t}
t (\lambda) = \mathrm{tr}_0 \left( K^+(\lambda) \mathcal{T}(\lambda) \right).
\end{equation}
The reflection matrix $K^+(\lambda)$ \eqref{K-plus} is the corresponding solution of the dual reflection equation \eqref{dRE}. The commutativity of the transfer matrix for different values of the spectral parameter
\begin{equation}
\label{open-tt}
[t (\lambda) , t (\mu)] = 0,
\end{equation}
is guaranteed by the dual reflection equation \eqref{dRE} and the exchange relations \eqref{exchangeRE} of the monodromy matrix $\mathcal{T}(\lambda)$ \cite{Sklyanin88}.

\section{Algebraic Bethe Ansatz}

In \cite{Eric13} it was shown that the most general case in which the algebraic Bethe ansatz can be fully implemented is when both K-matrices have upper-triangular from \eqref{K-minus} and \eqref{K-plus}. The main aim of this section is to define the Bethe vectors as to obtain the most simplest formulae for the off shell action of the transfer matrix of the spin chain on these Bethe vectors. The first step in this direction is to get the expressions of the entries of the monodromy matrix $\mathcal{T}(\lambda)$ in terms of the corresponding ones of the monodromies $T(\lambda)$ and  $\widetilde{T}(\lambda)$. According to definition of the monodromy matrix \eqref{calT} we have
\begin{equation}
\label{calT-tri}
\mathcal{T}(\lambda) = \left(\begin{array}{cc}
\mathcal{A} (\lambda) & \mathcal{B} (\lambda)  \\ 
\mathcal{C} (\lambda) & \mathcal{D} (\lambda)  
\end{array} \right) =
\left(\begin{array}{cc} 
A(\lambda ) &  B(\lambda ) \\ 
C(\lambda ) &  D(\lambda ) 
\end{array}\right) 
\left(\begin{array}{cc}
\xi ^{-} - \lambda \nu ^-  & \psi ^{-} \lambda \\ 
0 & \xi ^{-} + \lambda \nu ^- 
\end{array}\right)
\left(\begin{array}{cc} 
\widetilde{A}(\lambda ) &  \widetilde{B}(\lambda ) \\ 
\widetilde{C}(\lambda ) &  \widetilde{D}(\lambda ) 
\end{array}\right)  .
\end{equation}
%
%&= \left(\begin{array}{cc} 
%
%&  B(\lambda ) \\ 
%C(\lambda ) &  D(\lambda ) 
%\end{array}\right) 
From the equation above, using \eqref{T-til} and the RTT-relations \eqref{tTRT}, we obtain
\begin{align}
\label{CalA}
\mathcal{A} (\lambda) &=  (\xi ^{-} - \lambda \nu ^-) A(\lambda) \widetilde{A}(\lambda) + \left( (\psi ^{-} \lambda ) A(\lambda) + (\xi ^{-} + \lambda \nu ^-) B(\lambda) \right) \widetilde{C}(\lambda)  \\[1ex]
\label{CalD}
\mathcal{D} (\lambda) &=  (\xi ^{-} - \lambda \nu ^-) \left( \widetilde{B}(\lambda) C(\lambda ) 
- \frac{\eta}{2\lambda + \eta} \left( D(\lambda ) \widetilde{D}(\lambda) - \widetilde{A}(\lambda ) A(\lambda ) \right) \right) \notag \\
&+ \left( (\psi ^{-} \lambda ) C (\lambda) + (\xi ^{-} + \lambda \nu ^-) D(\lambda) \right) \widetilde{D}(\lambda) \\[1ex]
\label{CalB} 
\mathcal{B} (\lambda) &= (\xi ^{-} - \lambda \nu ^-) \left( \frac{2\lambda}{2\lambda + \eta} \widetilde{B}(\lambda ) A(\lambda) - \frac{\eta}{2\lambda + \eta} B(\lambda ) \widetilde{D}(\lambda ) \right)
+ \left( (\psi ^{-} \lambda ) A (\lambda) + (\xi ^{-} + \lambda \nu ^-) B(\lambda) \right) \widetilde{D}(\lambda) \\
\label{CalC}
\mathcal{C} (\lambda)&=  (\xi ^{-} - \lambda \nu ^-) C(\lambda) \widetilde{A}(\lambda) + \left( (\psi ^{-} \lambda ) C(\lambda) + (\xi ^{-} + \lambda \nu ^-) D(\lambda) \right) \widetilde{C}(\lambda)  .
\end{align}
With the aim of obtaining the action of the operators $\mathcal{A} (\lambda)$, $\mathcal{D} (\lambda)$ and $\mathcal{C} (\lambda)$ on the vector $\Omega _+$ \eqref{Omega+} we first observe that the action of the operators $\widetilde{A}(\lambda )$, $\widetilde{D}(\lambda)$ and $\widetilde{C}(\lambda)$ on the vector $\Omega _+$
\begin{eqnarray}
\label{AtilonOm}
\widetilde{A}(\lambda ) \Omega_+ &=& \widetilde{a} (\lambda) \Omega_+ , \quad \text{with}  \quad  \widetilde{a}(\lambda) = \prod _{m=1}^N \frac{\lambda + \alpha _m + \eta + \eta s_m}{\lambda + \alpha _m + \eta} , \\
\label{DtilonOm}    
\widetilde{D}(\lambda) \Omega_+  &=&  \widetilde{d}(\lambda) \Omega_+ , \quad \text{with}  \quad  \widetilde{d}(\lambda) = \prod _{m=1}^N \frac{\lambda + \alpha _m + \eta - \eta s_m}{\lambda + \alpha _m + \eta} , \\
\label{CtilonOm}
\widetilde{C}(\lambda) \Omega_+  &=& 0,  
\end{eqnarray}
follows directly from the definition \eqref{T-til}. Using the relations \eqref{CalA}-\eqref{CalC} and the formulas \eqref{AonOm}-\eqref{ConOm} and \eqref{AtilonOm}-\eqref{CtilonOm} we derive
\begin{eqnarray}
\label{CalConOm}
\mathcal{C} (\lambda) \Omega_+ &=& 0, \\
\label{CalAonOm}
\mathcal{A} (\lambda) \Omega_+ &=& \alpha (\lambda) \Omega_+ , \quad \text{with}  \quad \alpha (\lambda) =  (\xi ^{-} - \lambda \nu ^-) a(\lambda) \widetilde{a}(\lambda) , \\
\label{CalDonOm}
\mathcal{D} (\lambda) \Omega_+ &=& \delta (\lambda) \Omega_+ , \quad \text{with}  \\
\delta (\lambda) &=& \left( (\xi ^{-} + \lambda \nu ^-) - \frac{\eta}{2\lambda+\eta} (\xi ^{-} - \lambda \nu ^-) \right) d(\lambda) \widetilde{d}(\lambda) + \frac{\eta}{2\lambda+\eta} (\xi ^{-} - \lambda \nu ^-)  a(\lambda) \widetilde{a}(\lambda) .
\notag
\end{eqnarray}
In what follows we will use the fact that $\Omega_+$ is an eigenvector of the operator $\widehat{\mathcal{D}} (\lambda)$ \eqref{D-hat} 
\begin{equation}
\label{DhatOm}
\widehat{\mathcal{D}} (\lambda) \Omega_+ = \widehat{\delta} (\lambda) \Omega_+ , \quad \text{with}  \quad \widehat{\delta} (\lambda) = \delta (\lambda) - \frac{\eta}{2\lambda + \eta} \alpha (\lambda) ,
\end{equation}
or explicitly 
\begin{equation}
\label{deltahat}
\widehat{\delta }(\lambda) = \left( (\xi ^{-} + \lambda \nu ^-) - \frac{\eta}{2\lambda + \eta} (\xi ^{-} - \lambda \nu ^-) \right) d(\lambda) \widetilde{d}(\lambda) . %+ \frac{\eta ^2}{2\lambda (2\lambda + \eta)} (\xi ^{-} - \lambda \nu ^-)  a(\lambda) \widetilde{a}(\lambda) .
\end{equation}
The transfer matrix of the inhomogeneous XXX chain \eqref{open-t} with the triangular K-matrix \eqref{K-plus} can be expressed 
%\begin{align}
%\label{trans-mat}
%    t(\lambda) &= \mathrm{tr} \ K^{+}(\lambda) \mathcal{T}(\lambda) = \mathrm{tr}  \left(\begin{array}{cc}
%\xi  + \lambda + \eta & - \psi  (\lambda + \eta) \\ 0 & \xi  - \lambda - \eta \end{array}\right)
%\left(\begin{array}{cc}
%\mathcal{A} (\lambda) & \mathcal{B} (\lambda)  \\ \mathcal{C} (\lambda) & \mathcal{D} (\lambda)  \end{array}\right)
%  \notag \\[2ex]
%    &= \mathrm{tr} \ \left(\begin{array}{cc}
%(\xi  + \lambda + \eta) \mathcal{A} (\lambda) - \psi  (\lambda + \eta) \mathcal{C} (\lambda) & 
%(\xi  + \lambda + \eta) \mathcal{B} (\lambda)  (\lambda) - \psi  (\lambda + \eta) \mathcal{D} (\lambda) \\ 
%(\xi  - \lambda - \eta) \mathcal{C} (\lambda) & (\xi  - \lambda - \eta) \mathcal{D} (\lambda)  \end{array}\right) .
%\end{align} 
%Thus,
%\begin{equation}
%\label{transfer-mat}
%t(\lambda) =  (\xi^+  + \lambda + \eta) \mathcal{A} (\lambda) - \psi^+  (\lambda + \eta) \mathcal{C} (\lambda) + (\xi^+  - \lambda - \eta) \mathcal{D} (\lambda).
%\end{equation}
using Sklyanin's $\widehat{\mathcal{D}} (\lambda)$ operator \eqref{D-hat} \cite{Eric13}
%the transfer matrix is given by
\begin{equation}
\label{transfer-matrix}
t(\lambda) = \kappa _1 (\lambda) \mathcal{A} (\lambda) + \kappa _2 (\lambda) \widehat{\mathcal{D}} (\lambda) + \kappa _{12} (\lambda) \mathcal{C} (\lambda),
\end{equation}
with
\begin{equation}
\label{k-s}
\kappa _1 (\lambda) = 2 ( \xi^+ + \lambda \nu ^+) \ \frac{\lambda + \eta}{2 \lambda + \eta} , \quad
\kappa _2 (\lambda) =   \xi^+ - (\lambda + \eta) \nu ^+, \quad
\kappa _{12} (\lambda) = - \psi^+  (\lambda + \eta) .
\end{equation}

Evidently the vector $\Omega _+$ \eqref{Omega+} is an eigenvector  of the transfer matrix
\begin{equation}
\label{t-on-Om+}
t(\lambda) \Omega _+ = \left( \kappa _1 (\lambda) \alpha (\lambda) + \kappa _2 (\lambda) \widehat{\delta} (\lambda) \right) \Omega _+ = \Lambda _0 (\lambda) \Omega _+ . 
\end{equation}
For simplicity we have suppressed the dependence of the eigenvalue $\Lambda _0 (\lambda)$ on the boundary parameters $\xi ^+$ and $\nu ^+$ as well as the quasi-classical parameter $\eta$. 

We proceed to define the Bethe vectors $\Psi _M ( \mu_1 ,  \mu_2 ,  \dots ,  \mu_M )$ as to make the off shell action of $t(\lambda)$ on them as simple as possible. Before discussing $\Psi _M ( \mu_1 ,  \mu_2 ,  \dots ,  \mu_M )$, for arbitrary positive integer $M$, we will give explicitly first four Bethe vectors as well as the corresponding formulae for the off shell action of the transfer matrix. To this end, our next step is to show that 
\begin{equation}
\label{Psi1}
\Psi _1 (\mu) = \mathcal{B} (\mu) \Omega _+ + b_ 1 (\mu) \Omega _+ ,
\end{equation}
is a Bethe vector, if $b_ 1 (\mu)$ is chosen to be
\begin{equation}
\label{b1}
b_ 1 (\mu) = \frac{\psi^+}{2 \nu ^+} \left(  \frac{ 2\mu}{2\mu  + \eta}  \alpha (\mu ) -  \widehat{ \delta}  (\mu) \right) .
\end{equation}
A straightforward calculation, using the relations \eqref{comm-relAB}, \eqref{comm-rel-hDB} and \eqref{comm-relCB}, shows that off shell action of the transfer matrix \eqref{transfer-matrix} on $\Psi _1 (\mu)$ is given by
\begin{equation}
\label{t-on-Psi1}
\begin{split}
t(\lambda) \Psi _1 (\mu) = \Lambda _1 (\lambda , \mu ) \Psi _1 (\mu) + \frac{2 \eta (\lambda + \eta) ( \xi^+ + \mu \nu ^+)}{(\lambda - \mu) (\lambda + \mu + \eta)} F_1 (\mu) \Psi _1 (\lambda) 
\end{split}
\end{equation} 
where the eigenvalue $\Lambda _1 (\lambda , \mu )$  
is given by
\begin{equation}
\label{Lambda1}
\Lambda _1 (\lambda , \mu ) =  \kappa _1 (\lambda) \frac{(\lambda + \mu)(\lambda - \mu - \eta)}{(\lambda - \mu)(\lambda + \mu + \eta)} \alpha (\lambda) + \kappa _2 (\lambda) \frac{(\lambda - \mu + \eta)(\lambda + \mu + 2 \eta)}{(\lambda - \mu)(\lambda + \mu + \eta)} \widehat{\delta} (\lambda) .
\end{equation}
Evidently $\Lambda _1 (\lambda , \mu )$ depends also on boundary parameters $\xi ^+$, $\nu ^+$ and the quasi-classical parameter $\eta$, but these parameters are omitted in order to simplify the formulae. The unwanted term on the right hand side \eqref{t-on-Psi1} is annihilated by the Bethe equation 
\begin{equation}
\label{F1}
F_1 (\mu) = \frac{ 2\mu}{2\mu  + \eta}  \alpha (\mu ) - \frac{\xi^+ - (\mu + \eta)\nu ^+}{\xi^+ + \mu \nu ^+}  \widehat{ \delta}  (\mu) = 0,
\end{equation}
or equivalently,
\begin{equation}
\label{Bethe Eq-1}
\frac{\alpha (\mu)}{\widehat{ \delta} (\mu)} = \frac{(\mu  + \eta)\kappa _2 (\mu)}{\mu \, \kappa _1 (\mu)}
= \frac{(2 \mu  + \eta)(\xi^+ - (\mu + \eta)\nu ^+)}{2 \mu \, (\xi^+ + \mu \nu ^+)} .
\end{equation}
Therefore we have shown that $\Psi _1 (\mu)$ \eqref{Psi1} is the Bethe vector of the transfer matrix \eqref{transfer-matrix} corresponding to the eigenvalue $\Lambda _1 (\lambda , \mu )$ \eqref{Lambda1}.

We seek the Bethe vector $\Psi _2 ( \mu_1 ,  \mu_2)$ in the form 
\begin{equation}
\label{Psi2}
\Psi _2 ( \mu_1 ,  \mu_2) = \mathcal{B} (\mu_1) \mathcal{B} (\mu_2) \Omega _+ +  b^{(1)}_2(\mu_2 ; \mu_1)  \mathcal{B} (\mu_1) \Omega _+  +  b^{(1)}_2(\mu_1; \mu_2 ) \mathcal{B} (\mu_2) \Omega _+ +  b^{(2)}_2 ( \mu_1 ,  \mu_2) \Omega _+ ,
\end{equation}
where $b^{(1)}_2(\mu_1; \mu_2 )$ and $b^{(2)}_2( \mu_1 ,  \mu_2)$ are given by
\begin{align}
\label{1b2}
b^{(1)}_2 (\mu_1; \mu_2 ) &= \frac{\psi^+}{2 \nu ^+} \left( \frac{ 2 \mu _1}{2 \mu _1  + \eta}
\frac{(\mu _1 + \mu _2)(\mu _1 - \mu _2 - \eta )}{(\mu _1 - \mu _2)(\mu _1 + \mu _2 + \eta )} \alpha (\mu _1) 
- \frac{(\mu _1 - \mu_2 + \eta)(\mu _1 + \mu_2 + 2 \eta)}{(\mu _1 - \mu _2)(\mu _1 + \mu_2 + \eta)} \widehat{ \delta}  (\mu _1) \right) ,  \\[2ex]
\label{2b2}
b^{(2)}_2 ( \mu_1 ,  \mu_2) &=  \frac{1}{2} \left(  b^{(1)}_2 (\mu_1; \mu_2 ) \, b_1 (\mu _2) +
b^{(1)}_2 (\mu_2; \mu_1 )  \, b_1 (\mu _1)  \right) .
\end{align}

Starting from the definitions \eqref{transfer-matrix} and \eqref{Psi2}, using the relations \eqref{ABBOmega}, \eqref{DBBOmega} and \eqref{CBBOmega} to push the operators $\mathcal{A} (\lambda)$, $\widehat{\mathcal{D}} (\lambda)$ and $\mathcal{C} (\lambda)$ to the right and after rearranging some terms, we obtain the off shell action of transfer matrix $t(\lambda)$ on $\Psi _2 ( \mu_1 ,  \mu_2)$
\begin{align}
\label{t-on-Psi2}
&t(\lambda) \Psi _2 ( \mu_1 ,  \mu_2) = \Lambda _2 (\lambda , \{\mu _i \} ) \Psi _2 ( \mu_1 ,  \mu_2) 
+ \sum _{i=1}^2 \frac{2 \eta (\lambda + \eta ) (\xi^+ + \mu _i \nu ^+)}{(\lambda - \mu _i) (\lambda + \mu _i + \eta )} F_2(\mu _i; \mu _{3-i}) \Psi _2 ( \lambda ,  \mu_{3-i}) ,
\end{align}
where the eigenvalue is given by
\begin{equation}
\label{Lambda2} 
\Lambda _2 (\lambda , \{\mu _i \}) = \kappa _1 (\lambda) \ \alpha (\lambda) \ \prod _{i =1}^2
\frac{(\lambda + \mu_i)(\lambda - \mu_i - \eta)}{(\lambda - \mu _i)(\lambda + \mu_i + \eta)}  
+ \kappa _2 (\lambda) \ \widehat{ \delta} (\lambda) \ \prod _{i =1}^2
\frac{(\lambda - \mu_i + \eta)(\lambda + \mu_i + 2 \eta )}{(\lambda - \mu_i )(\lambda + \mu _i + \eta)} 
\end{equation}
and the two unwanted terms in \eqref{t-on-Psi2} are canceled by the Bethe equations which follow from $F_2(\mu _i; \mu _{3-i}) = 0$, i. e.
\begin{equation}
\label{BE-2}
\frac{ 2\mu _i}{2 \mu _i  + \eta} \
\frac{(\mu _ i + \mu _{3-i})(\mu _ i - \mu _{3-i} - \eta )}{(\mu _ i - \mu _{3-i})(\mu _ i + \mu _{3-i} + \eta )} \alpha (\mu _i) 
- \frac{\xi^+ - (\mu _i + \eta)\nu ^+}{\xi^+ + \mu _i \nu ^+} \ \frac{(\mu _i - \mu_{3-i} + \eta)(\mu _i + \mu_{3-i} + 2 \eta)}{(\mu _i - \mu _{3-i})(\mu _i + \mu_{3-i} + \eta)} \widehat{ \delta}  (\mu _i) = 0 ,  %\\
\end{equation}
with $i = \{ 1, 2 \}$. 
Therefore the Bethe equations are
\begin{equation}
\label{Bethe Eq-2.i}
\frac{\alpha (\mu _i)}{\widehat{ \delta} (\mu _i)} = \frac{(\mu _i  + \eta)\kappa _2 (\mu _i)}{\mu _i \, \kappa _1 (\mu _i)} \frac{(\mu _i - \mu_{3-i} + \eta)(\mu _i + \mu_{3-i} + 2 \eta)}{(\mu _ i + \mu _{3-i})(\mu _ i - \mu _{3-i} - \eta )} ,
\end{equation}
where $i = \{ 1, 2 \}$. Striking property of the Bethe vectors we have introduced so far is the simplicity of the off shell action of the transfer matrix $t(\lambda)$, equations \eqref{t-on-Psi1} and  \eqref{t-on-Psi2}.  Actually, the action of the transfer matrix almost coincides with the one in the case when the two boundary matrices are diagonal \cite{Sklyanin88, Hikami95}.

The Bethe vector $\Psi _3 ( \mu_1 ,  \mu_2 ,  \mu_3)$ we propose is a symmetric function of its arguments and it is given as the following sum of eight terms
\begin{equation}
\label{Psi3}
\begin{split}
\Psi _3 ( \mu_1 ,  \mu_2 ,  \mu_3) = \mathcal{B} (\mu_1) \mathcal{B} (\mu_2) \mathcal{B} (\mu_3)\Omega _+ 
+  b ^{(1)}_3(\mu_3 ; \mu_2 , \mu_1) \mathcal{B} (\mu_1) \mathcal{B} (\mu_2) \Omega _+ 
+ b ^{(1)}_3(\mu_1 ;  \mu_2 ,  \mu_3)  \times \\
\times \mathcal{B} (\mu_2) \mathcal{B} (\mu_3) \Omega _+ 
+ b ^{(1)}_3(\mu_2 ; \mu_1 , \mu_3) \mathcal{B} (\mu_1) \mathcal{B} (\mu_3) \Omega _+ 
+ b ^{(2)}_3(\mu_1 ,  \mu_2 ;  \mu_3)  \mathcal{B} (\mu_3) \Omega _+  \\
+ b ^{(2)}_3(\mu_1, \mu_3 ;  \mu_2) \mathcal{B} (\mu_2) \Omega _+ 
+ b ^{(2)}_3(\mu_2, \mu_3 ;  \mu_1) \mathcal{B} (\mu_2) \Omega _+ 
+ b ^{(3)}_3 ( \mu_1 ,  \mu_2 ,  \mu_3) \Omega _+ ,
\end{split}
\end{equation}
where $b ^{(1)}_3 (\mu_1 ;  \mu_2 ,  \mu_3)$ , $b ^{(2)}_3 (\mu_1 ,  \mu_2 ;  \mu_3)$ and $b ^{(3)}_3 (\mu_1 ,  \mu_2 ,  \mu_3)$ are given by
\begin{align}
\label{b3-1}
b ^{(1)}_3 (\mu_1 ;  \mu_2 ,  \mu_3) &=  \frac{\psi^+}{2 \nu ^+} \left( \frac{ 2 \mu _1}{2 \mu _1  + \eta}
\alpha (\mu _1) \ \prod _{j =2}^3
\frac{(\mu _1 + \mu _j)(\mu _1 - \mu _j - \eta )}{(\mu _1 - \mu _j)(\mu _1 + \mu _j + \eta )} 
\right. \notag \\[1ex]
&\left. -  \widehat{ \delta}  (\mu _1) \ \prod _{j =2}^3 \frac{(\mu _1 - \mu_j + \eta)(\mu _1 + \mu_j + 2 \eta)}{(\mu _1 - \mu _j)(\mu _1 + \mu_j + \eta)}  \right) ,  \\[1ex]
\label{b3-2}
b ^{(2)}_3 (\mu_1 ,  \mu_2 ;  \mu_3) &= \frac{1}{2} \left( b^{(1)}_3 (\mu_1; \mu_2 , \mu_3 )  \, b^{(1)}_2 (\mu_2; \mu_3 ) +b^{(1)}_3 (\mu_2; \mu_1 , \mu_3 )  \, b^{(1)}_2 (\mu_1; \mu_3 ) \right) ,  \\[1ex]
\label{b3-3}
b ^{(3)}_3 (\mu_1 ,  \mu_2 ,  \mu_3) &=  \frac{1}{6} \left(  b^{(1)}_3 (\mu_1; \mu_2 , \mu_3 )  \, b^{(1)}_2 (\mu_2; \mu_3 ) \, b_1 (\mu _3) +  b^{(1)}_3 (\mu_1; \mu_2 , \mu_3 )  \, b^{(1)}_2 (\mu_3 ; \mu_2 ) \, b_1 (\mu _2) \right. \notag \\[1ex]
&\left. + b^{(1)}_3 (\mu_2; \mu_1 , \mu_3 )  \, b^{(1)}_2 (\mu_1 ; \mu_3 ) \, b_1 (\mu _3)
+ b^{(1)}_3 (\mu_2; \mu_1 , \mu_3 )  \, b^{(1)}_2 (\mu_3 ; \mu_1 ) \, b_1 (\mu _1) \right. \notag \\[1ex]
&\left. + b^{(1)}_3 (\mu_3 ; \mu_1 , \mu_2 )  \, b^{(1)}_2 (\mu_1 ; \mu_2 ) \, b_1 (\mu _2)
+ b^{(1)}_3 (\mu_3; \mu_1 , \mu_2 )  \, b^{(1)}_2 (\mu_2 ; \mu_1 ) \, b_1 (\mu _1) \right) .
\end{align}
The action of $t(\lambda)$ \eqref{transfer-matrix} on $\Psi _3 ( \mu_1 ,  \mu_2 ,  \mu_3)$, obtained using evident generalization of the formulas \eqref{ABBOmega}, \eqref{DBBOmega} and  \eqref{CBBOmega} and subsequent rearranging of terms, reads
\begin{equation}
\label{t-on-Psi3}
\begin{split} 
t(\lambda) \Psi _3 ( \mu_1 ,  \mu_2 ,  \mu_3) = \Lambda _3 (\lambda , \{\mu _i \}) \Psi _3 ( \mu_1 ,  \mu_2 ,  \mu_3) + \sum _{i=1}^3 \frac{2 \eta (\lambda + \eta ) (\xi^+ + \mu _i \nu ^+)}{(\lambda - \mu _i) (\lambda + \mu _i + \eta )} F_3 (\mu _i ; \{\mu _j \} _{j \neq i}) \Psi _3 ( \lambda ,  \{\mu _j \} _{j \neq i}) 
\end{split}
\end{equation}
where the eigenvalue is given by
\begin{equation}
\label{Lambda3}
\begin{split} 
\Lambda _3 (\lambda , \{\mu _i \}) = \kappa _1 (\lambda) \ \alpha (\lambda) \ \prod _{i =1}^3
\frac{(\lambda + \mu_i)(\lambda - \mu_i - \eta)}{(\lambda - \mu _i)(\lambda + \mu_i + \eta)}  
+ \kappa _2 (\lambda) \ \widehat{ \delta} (\lambda) \ \prod _{i =1}^3
\frac{(\lambda - \mu_i + \eta)(\lambda + \mu_i + 2 \eta )}{(\lambda - \mu_i )(\lambda + \mu _i + \eta)} 
\end{split}
\end{equation}
and the three unwanted terms o the right hand side of \eqref{t-on-Psi3} are canceled by the Bethe equations $F_3(\mu _i ; \{\mu _j \} _{j \neq i}) = 0$, explicitly
\begin{equation}
\label{BE-3}
\begin{split} 
\frac{ 2\mu _i}{2 \mu _i  + \eta} \
\alpha (\mu _i) \ \prod _{\substack{ j =1\\ j \neq i}}^3
\frac{(\mu _ i + \mu _j)(\mu _ i - \mu _j - \eta )}{(\mu _ i - \mu _j)(\mu _i + \mu _j + \eta )} 
- \frac{\xi^+ - (\mu _i + \eta)\nu ^+}{\xi^+ + \mu _i \nu ^+} \ \widehat{ \delta}  (\mu _i) \ 
\prod _{\substack{ j =1\\ j \neq i}}^3 \frac{(\mu _i - \mu_j + \eta)(\mu _i + \mu_j + 2 \eta)}{(\mu _i - \mu _j)(\mu _i + \mu_j + \eta)} 
= 0 , 
\end{split}
\end{equation}
with $i = \{ 1 , 2 , 3 \}$, or in another form
\begin{equation}
\label{Bethe Eq-3.2}
\frac{\alpha (\mu _i)}{\widehat{ \delta} (\mu _i)} =  \frac{(\mu _i  + \eta)\kappa _2 (\mu _i)}{\mu _i \, \kappa _1 (\mu _i)} \ \prod _{\substack{ j =1\\ j \neq i}}^3 \frac{(\mu _i - \mu_j+ \eta)(\mu _i + \mu_j + 2 \eta)}{(\mu _i + \mu _j)(\mu _i - \mu _j - \eta )}, 
\end{equation}
for $i = \{ 1 , 2 , 3 \}$. As it is evident from \eqref{t-on-Psi3}, our choice of the Bethe vector $\Psi _3 ( \mu_1 ,  \mu_2 ,  \mu_3)$ \eqref{Psi3} makes the off shell action of the transfer matrix strikingly simple.

With the aim of making the presentation more transparent, still before addressing the general $\Psi _M ( \mu_1 ,  \mu_2 ,  \dots ,  \mu_M )$, we show explicit formulas for $\Psi _4 ( \mu_1 ,  \mu_2 ,  \mu_3 ,  \mu_4 )$. The Bethe vector $\Psi _4 ( \mu_1 ,  \mu_2 ,  \mu_3 ,  \mu_4 )$ is a symmetric function of its arguments and as a sum of sixteen terms it reads
\begin{equation}
\label{Psi4}
\begin{split}
&\Psi _4 ( \mu_1 ,  \mu_2 ,  \mu_3 ,  \mu_4 ) = \mathcal{B} (\mu_1) \mathcal{B} (\mu_2) \mathcal{B} (\mu_3) \mathcal{B} (\mu_4) \Omega _+ 
+  b ^{(1)}_4(\mu_4 ; \mu_1, \mu_2 , \mu_3) \mathcal{B} (\mu_1) \mathcal{B} (\mu_2) \mathcal{B} (\mu_3) \Omega _+ 
 \\
&+ b ^{(1)}_4(\mu_3 ; \mu_1 , \mu_2 , \mu_4) \mathcal{B} (\mu_1) \mathcal{B} (\mu_2) \mathcal{B} (\mu_4) \Omega _+
+ b ^{(1)}_4(\mu_2 ; \mu_1 , \mu_3 , \mu_4) \mathcal{B} (\mu_1) \mathcal{B} (\mu_3) \mathcal{B} (\mu_4) \Omega _+  \\
&+ b ^{(1)}_4(\mu_1 ;  \mu_2 ,  \mu_3 ,  \mu_4) \mathcal{B} (\mu_2) \mathcal{B} (\mu_3) \mathcal{B} (\mu_4) \Omega _+ 
+b ^{(2)}_4(\mu_3 ,  \mu_4 ;  \mu_1 ,  \mu_2)  \mathcal{B} (\mu_1) \mathcal{B} (\mu_2) \Omega _+ \\
&+b ^{(2)}_4(\mu_2 ,  \mu_4 ;  \mu_1 ,  \mu_3)  \mathcal{B} (\mu_1) \mathcal{B} (\mu_3) \Omega _+
+b ^{(2)}_4(\mu_2 ,  \mu_3 ;  \mu_1 ,  \mu_4)  \mathcal{B} (\mu_1) \mathcal{B} (\mu_4) \Omega _+  \\
&+b ^{(2)}_4(\mu_1 ,  \mu_4 ;  \mu_2 ,  \mu_3)  \mathcal{B} (\mu_2) \mathcal{B} (\mu_3) \Omega _+
+b ^{(2)}_4(\mu_1 ,  \mu_3 ;  \mu_2 ,  \mu_4)  \mathcal{B} (\mu_2) \mathcal{B} (\mu_4) \Omega _+  \\
&+ b ^{(2)}_4(\mu_1 ,  \mu_1 ;  \mu_3 ,  \mu_4)  \mathcal{B} (\mu_3) \mathcal{B} (\mu_4) \Omega _+  
+ b ^{(3)}_4(\mu_2, \mu_3  ,  \mu_4 ;  \mu_1) \mathcal{B} (\mu_1) \Omega _+ 
+ b ^{(3)}_4(\mu_1, \mu_3  ,  \mu_4 ;  \mu_2) \mathcal{B} (\mu_2) \Omega _+ 
\\
&+ b ^{(3)}_4(\mu_1, \mu_2  ,  \mu_4 ;  \mu_3) \mathcal{B} (\mu_3) \Omega _+ 
+ b ^{(3)}_4(\mu_1, \mu_2  ,  \mu_3 ;  \mu_4) \mathcal{B} (\mu_4)  \Omega _+ 
+ b ^{(4)}_4 ( \mu_1 ,  \mu_2 ,  \mu_3 ,  \mu_4 ) \Omega _+ ,
\end{split}
\end{equation}
where the coefficients are given by
\begin{align}
\label{b4-1}
 b ^{(1)}_4(\mu_1 ;  \mu_2 ,  \mu_3 ,  \mu_4) &= \frac{\psi^+}{2 \nu ^+} \left( \frac{ 2 \mu _1}{2 \mu _1  + \eta}
\alpha (\mu _1) \ \prod _{j =2}^4
\frac{(\mu _1 + \mu _j)(\mu _1 - \mu _j - \eta )}{(\mu _1 - \mu _j)(\mu _1 + \mu _j + \eta )} 
\right. \notag \\[1ex]
&\left. -  \widehat{ \delta}  (\mu _1) \ \prod _{j =2}^4 \frac{(\mu _1 - \mu_j + \eta)(\mu _1 + \mu_j + 2 \eta)}{(\mu _1 - \mu _j)(\mu _1 + \mu_j + \eta)}  \right) ,  \\[1ex]
\label{b4-2}
b ^{(2)}_4(\mu_1 ,  \mu_2 ;  \mu_3 ,  \mu_4) &= \frac{1}{2} \left(  b^{(1)}_4 (\mu_1; \mu_2 , \mu_3 ,  \mu_4 )  \, b^{(1)}_3 (\mu_2; \mu_3 , \mu_4) +b^{(1)}_4 (\mu_2; \mu_1 , \mu_3 ,  \mu_4 )  \, b^{(1)}_3 (\mu_1; \mu_3 , \mu_4) \right) , 
\end{align}

\begin{align}
\label{b4-3}
b ^{(3)}_4(\mu_1, \mu_2  ,  \mu_3 ;  \mu_4) &=  \frac{1}{3!} \sum_{\rho \in S_3} b^{(1)}_4 (\mu_{\rho (1)}; \mu_{\rho (2)} , \mu_{\rho (3)} , \mu_{4} )  \, b^{(1)}_3 (\mu_{\rho (2)} ;  \mu_{\rho (3)} , \mu_{4}) \, b^{(1)}_2 ( \mu_{\rho (3)} ; \mu_{4}) \\[1ex]
\label{b4-4}
b ^{(4)}_4 ( \mu_1 ,  \mu_2 ,  \mu_3 ,  \mu_4 ) &=   \frac{1}{4!} \sum_{\sigma \in S_4} b^{(1)}_4 (\mu_{\sigma (1)}; \mu_{\sigma (2)} , \mu_{\sigma (3)} , \mu_{\sigma (4)} )  \, b^{(1)}_3 (\mu_{\sigma (2)} ;  \mu_{\sigma (3)} , \mu_{\sigma (4)}) \, b^{(1)}_2 ( \mu_{\sigma (3)} ;\mu_{\sigma (4)}) \, b_1 (\mu_{\sigma (4)}) ,
\end{align}
where $S_3$ and $S_4$ are the symmetric groups of degree 3 and 4, respectively. An analogous calculation to the one in the previous case \eqref{t-on-Psi3}, just bit longer, shows that the off shell action of the transfer matrix \eqref{transfer-matrix} on $\Psi _4 ( \mu_1 ,  \mu_2 ,  \mu_3, \mu_4)$ is given by
\begin{equation}
\label{t-on-Psi4}
\begin{split} 
t(\lambda) \Psi _4 ( \mu_1 ,  \mu_2 ,  \mu_3, \mu_4) = \Lambda _4 (\lambda , \{\mu _i \}) \Psi _4 ( \mu_1 ,  \mu_2 ,  \mu_3, \mu_4)  \\[1ex]
+ \sum _{i=1}^4 \frac{2 \eta (\lambda + \eta ) (\xi^+ + \mu _i \nu ^+)}{(\lambda - \mu _i) (\lambda + \mu _i + \eta )} F_4 (\mu _i ; \{\mu _j \} _{j \neq i}) \Psi _4 ( \lambda ,  \{\mu _j \} _{j \neq i}) 
\end{split}
\end{equation}
where
\begin{equation}
\label{Lambda4}
\begin{split} 
\Lambda _4 (\lambda , \{\mu _i \}) = \kappa _1 (\lambda) \ \alpha (\lambda) \ \prod _{i =1}^4
\frac{(\lambda + \mu_i)(\lambda - \mu_i - \eta)}{(\lambda - \mu _i)(\lambda + \mu_i + \eta)}  
+ \kappa _2 (\lambda) \ \widehat{ \delta} (\lambda) \ \prod _{i =1}^4
\frac{(\lambda - \mu_i + \eta)(\lambda + \mu_i + 2 \eta )}{(\lambda - \mu_i )(\lambda + \mu _i + \eta)} 
\end{split}
\end{equation}
and the four unwanted terms o the right hand side of \eqref{t-on-Psi3} are canceled by the four Bethe equations $F_4(\mu _i ; \{\mu _j \} _{j \neq i}) = 0$, explicitly
\begin{equation}
\label{BE-4}
\begin{split} 
\frac{ 2\mu _i}{2 \mu _i  + \eta} \
\alpha (\mu _i) \ \prod _{\substack{ j =1\\ j \neq i}}^4
\frac{(\mu _ i + \mu _j)(\mu _ i - \mu _j - \eta )}{(\mu _ i - \mu _j)(\mu _i + \mu _j + \eta )} 
- \frac{\xi^+ - (\mu _i + \eta)\nu ^+}{\xi^+ + \mu _i \nu ^+} \ \widehat{ \delta}  (\mu _i) \ 
\prod _{\substack{ j =1\\ j \neq i}}^4 \frac{(\mu _i - \mu_j + \eta)(\mu _i + \mu_j + 2 \eta)}{(\mu _i - \mu _j)(\mu _i + \mu_j + \eta)} 
= 0 , 
\end{split}
\end{equation}
or equivalently 
\begin{equation}
\label{Bethe Eq-3.2}
\frac{\alpha (\mu _i)}{\widehat{ \delta} (\mu _i)} =  \frac{(\mu _i  + \eta)\kappa _2 (\mu _i)}{\mu _i \, \kappa _1 (\mu _i)} \ \prod _{\substack{ j =1\\ j \neq i}}^4 \frac{(\mu _i - \mu_j+ \eta)(\mu _i + \mu_j + 2 \eta)}{(\mu _i + \mu _j)(\mu _i - \mu _j - \eta )}, 
\end{equation}
with $i = \{ 1 , 2 , 3 , 4 \}$. 

We proceed to define $\Psi _M ( \mu_1 ,  \mu_2 ,  \dots ,  \mu_M )$ as a sum of $2^M$ terms, for arbitrary positive integer $M$, and as a symmetric function of its arguments 
\begin{equation}
\label{PsiM}
\begin{split}
&\Psi _M ( \mu_1 ,  \mu_2 ,  \dots ,  \mu_M ) = \mathcal{B} (\mu_1) \mathcal{B} (\mu_2) \cdots \mathcal{B} (\mu_M) \Omega _+ 
+  b ^{(1)}_M(\mu_M ; \mu_1 ,  \mu_2 ,  \dots ,  \mu_{M-1})\mathcal{B} (\mu_1) \mathcal{B} (\mu_2) \cdots \mathcal{B} (\mu_{M-1}) \Omega _+  \\
&+\cdots + b ^{(2)}_M(\mu_{M-1} , \mu_M  ; \mu_1 ,  \mu_2 ,  \dots ,  \mu_{M-2}) \mathcal{B} (\mu_1) \mathcal{B} (\mu_2) \cdots \mathcal{B} (\mu_{M-2}) \Omega _+   \\
&\ \ \vdots   \\
&+ b ^{(M-1)}_M (\mu_1 ,  \mu_2 ,  \dots ,  \mu_{M-1} ;  \mu_M) \mathcal{B} (\mu_M)  \Omega _+
+ b ^{(M)}_M ( \mu_1 ,  \mu_2 ,  \dots ,  \mu_M) \Omega _+ ,
\end{split}
\end{equation}
where the coefficients are given by
\begin{align}
\label{bM-1}
 b ^{(1)}_M(\mu_1 ;  \mu_2 ,  \mu_3 ,  \dots ,  \mu_M) &= \frac{\psi^+}{2 \nu ^+} \left( \frac{ 2 \mu _1}{2 \mu _1  + \eta}
\alpha (\mu _1) \ \prod _{j =2}^M
\frac{(\mu _1 + \mu _j)(\mu _1 - \mu _j - \eta )}{(\mu _1 - \mu _j)(\mu _1 + \mu _j + \eta )} 
\right. \notag \\[1ex]
&\left. -  \widehat{ \delta}  (\mu _1) \ \prod _{j =2}^M \frac{(\mu _1 - \mu_j + \eta)(\mu _1 + \mu_j + 2 \eta)}{(\mu _1 - \mu _j)(\mu _1 + \mu_j + \eta)}  \right) ,  \\[2ex]
\label{bM-2}
b ^{(2)}_M(\mu_1 , \mu_2 ;  \mu_3 ,  \dots ,  \mu_M) &= \frac{1}{2} \left(  b ^{(1)}_M(\mu_1 ;  \mu_2 ,  \mu_3 ,  \dots ,  \mu_M) b ^{(1)}_{M-1}(\mu_2 ;  \mu_3 ,  \dots ,  \mu_M) \right. \notag \\[1ex]
&\left. +  b ^{(1)}_M(\mu_2 ;  \mu_1 ,  \mu_3 ,  \dots ,  \mu_M) b ^{(1)}_{M-1}(\mu_1 ;  \mu_3 ,  \dots ,  \mu_M)
\right) , \\
&\ \ \vdots \notag \\
\label{bM-(M-1)}
b ^{(M-1)}_M (\mu_1 ,  \mu_2 ,  \dots ,  \mu_{M-1} ;  \mu_M) &=  \frac{1}{(M-1)!} \sum_{\rho \in S_{M-1} } b^{(1)}_M (\mu_{\rho (1)}; \mu_{\rho (2)} , \dots , \mu_{M} )  \, b ^{(1)}_{M-1}(\mu_{\rho (2)} ;  \mu_{\rho (3)} ,  \dots ,  \mu_M) \times \notag \\[1ex]
&\times b ^{(1)}_{M-2}(\mu_{\rho (3)} ;  \mu_{\rho (4)} ,  \dots ,  \mu_M) \cdots b^{(1)}_2 ( \mu_{\rho (M-1)} ; \mu_{M}) \\[2ex]
\label{bM-M}
b ^{(M)}_M ( \mu_1 ,  \mu_2 ,  \dots ,  \mu_M) &=   \frac{1}{M!} \sum_{\sigma \in S_M} b^{(1)}_M (\mu_{\sigma (1)}; \mu_{\sigma (2)} , \dots , \mu_{\sigma (M)} )  \, b ^{(1)}_{M-1}(\mu_{\sigma (2)} ;  \mu_{\sigma (3)} ,  \dots ,  \mu_{\sigma (M)}) \times \notag \\[1ex]
&\times b ^{(1)}_{M-2}(\mu_{\sigma (3)} ;  \mu_{\sigma (4)} ,  \dots , \mu_{\sigma (M)}) \cdots b^{(1)}_2 ( \mu_{\sigma (M-1)} ;\mu_{\sigma (M)}) \, b_1 (\mu_{\sigma (M)}) ,
\end{align}
where $S_{M-1}$ and $S_M$ are the symmetric groups of degree M-1 and M, respectively. 

A straightforward calculation based on evident generalization of the formulas \eqref{ABBOmega}, \eqref{DBBOmega} and  \eqref{CBBOmega} and subsequent rearranging of terms, yields
the off shell action of the transfer matrix on the Bethe vector $\Psi _M ( \mu_1 ,  \mu_2 ,  \dots ,  \mu_M )$ 
\begin{equation}
\label{t-on-PsiM}
\begin{split} 
t(\lambda) \Psi _M ( \mu_1 ,  \mu_2 ,  \dots ,  \mu_M )  = \Lambda _M (\lambda , \{\mu _i \}) \Psi _M ( \mu_1 ,  \mu_2 ,  \dots ,  \mu_M )   \\[1ex]
+ \sum _{i=1}^M \frac{2 \eta (\lambda + \eta ) (\xi^+ + \mu _i \nu ^+)}{(\lambda - \mu _i) (\lambda + \mu _i + \eta )} F_M (\mu _i ; \{\mu _j \} _{j \neq i}) \Psi _M ( \lambda ,  \{\mu _j \} _{j \neq i}) ,
\end{split}
\end{equation}
where the corresponding eigenvalue is given by
\begin{equation}
\label{LambdaM}
\begin{split} 
\Lambda _M (\lambda , \{\mu _i \}) = \kappa _1 (\lambda) \ \alpha (\lambda) \ \prod _{i =1}^M
\frac{(\lambda + \mu_i)(\lambda - \mu_i - \eta)}{(\lambda - \mu _i)(\lambda + \mu_i + \eta)}  
+ \kappa _2 (\lambda) \ \widehat{ \delta} (\lambda) \ \prod _{i =1}^M
\frac{(\lambda - \mu_i + \eta)(\lambda + \mu_i + 2 \eta )}{(\lambda - \mu_i )(\lambda + \mu _i + \eta)} 
\end{split}
\end{equation}
and the $M$ unwanted terms o the right hand side of \eqref{t-on-Psi3} are canceled by the Bethe equations $F_M(\mu _i ; \{\mu _j \} _{j \neq i}) = 0$, explicitly
\begin{equation}
\label{BE-M}
\begin{split} 
\frac{ 2\mu _i}{2 \mu _i  + \eta} \
\alpha (\mu _i) \ \prod _{\substack{ j =1\\ j \neq i}}^M
\frac{(\mu _ i + \mu _j)(\mu _ i - \mu _j - \eta )}{(\mu _ i - \mu _j)(\mu _i + \mu _j + \eta )} 
- \frac{\xi^+ - (\mu _i + \eta)\nu ^+}{\xi^+ + \mu _i \nu ^+} \ \widehat{ \delta}  (\mu _i) \ 
\prod _{\substack{ j =1\\ j \neq i}}^M \frac{(\mu _i - \mu_j + \eta)(\mu _i + \mu_j + 2 \eta)}{(\mu _i - \mu _j)(\mu _i + \mu_j + \eta)} 
= 0 , 
\end{split}
\end{equation}
or equivalently 
\begin{equation}
\label{Bethe Eq-3.2}
\frac{\alpha (\mu _i)}{\widehat{ \delta} (\mu _i)} =  \frac{(\mu _i  + \eta)\kappa _2 (\mu _i)}{\mu _i \, \kappa _1 (\mu _i)} \ \prod _{\substack{ j =1\\ j \neq i}}^M \frac{(\mu _i - \mu_j+ \eta)(\mu _i + \mu_j + 2 \eta)}{(\mu _i + \mu _j)(\mu _i - \mu _j - \eta )}, 
\end{equation}
with $i = \{ 1 , 2 , \dots , M \}$. The Bethe vectors $\Psi _M ( \mu_1 ,  \mu_2 ,  \dots ,  \mu_M )$ we have defined in \eqref{PsiM} yield the strikingly simple expression \eqref{t-on-PsiM} for the off shell action of the transfer matrix $t(\lambda)$ \eqref{transfer-matrix}. Actually, the action of the transfer matrix is as simple as it could possible be since it almost coincides with the one in the case when the two boundary matrices are diagonal \cite{Sklyanin88, Hikami95}. In this way we have fully implemented the algebraic Bethe ansatz for the XXX spin chain in the case when both boundary matrices have upper-triangular form \eqref{K-minus} and \eqref{K-plus}.

\section{Gaudin Model}
We explore further the results obtained in the previous section on the XXX Heisenberg spin chain in the case when both boundary matrix are upper-triangular. We combine them together with the quasi-classical limit studied in \cite{CAMRS} with the aim of implementing fully the off shell Bethe ansatz for the corresponding Gaudin model by defining the Bethe vectors and deriving its spectrum and the corresponding Bethe equations.  

For the study of the open Gaudin model we impose
\begin{equation}
\label{normalizationKpl}
\lim_{\eta \to 0}\Big(  K^+(\lambda) K^{-} (\lambda)\Big)  = \left(\xi^2-\lambda ^2 \nu ^2 \right) \mathbbm{1}.
\end{equation}
In particular, this implies that the parameters of the reflection matrices on the left and on the right end of the chain are the same. In general, this is not the case in  the study of the open spin chain. However, this condition is essential for the Gaudin model. Then we will write
\begin{equation}
\label{K-min-GM}
K^-(\lambda)\equiv K(\lambda) = \left(
\begin{array}{cc}
\xi  - \lambda \nu & \lambda \psi \\
0 & \xi  + \lambda \nu 
\end{array} \right) ,
\end{equation}
so that 
\begin{equation}
\label{def-KplusGM}
K^+(\lambda)= K(-\lambda-\eta)= \left(
 \begin{array}{cc}
 \xi  + ( \lambda + \eta ) \nu &   - \psi  (\lambda + \eta) \\
 0 & \xi - ( \lambda + \eta ) \nu
 \end{array}
 \right).
\end{equation}

In \cite{CAMRS} we have derived the generating function of the Gaudin Hamiltonians with boundary terms following Sklyanin's approach in the periodic case \cite{Sklyanin89}. Our derivation is based on the quasi-classical expansion of the linear combination of the transfer matrix of the XXX chain and the central element, the so-called Sklyanin determinant. Finally, the expansion reads \cite{CAMRS} 
\begin{align}
\label{final-exp-t-D-open}
2 \lambda t(\lambda) - \Delta \left[\mathcal{T}(\lambda)\right]   &= 2 \lambda \left(\xi^2 - \lambda ^2 \nu ^2 \right) \mathbbm{1} + \eta \left(\xi^2 - 3 \lambda ^2 \nu ^2 \right)  \mathbbm{1} \notag\\
&+ \eta ^2  \lambda \left( \left( \xi^2 - \lambda ^2 \nu ^2 \right) \tau (\lambda) - \frac{\nu ^2}{2} \mathbbm{1} \right) + \mathcal{O}(\eta ^3) ,
\end{align}
where $\tau (\lambda)$ is the generating function of the Gaudin Hamiltonians, with upper triangular reflection matrix \eqref{K-min-GM},
\begin{equation}
\label{open-tau} 
\tau (\lambda) =  \mathrm{tr}_0 \, \mathcal{L}_0 ^2(\lambda) ,
\end{equation} 
and the Lax matrix 
\begin{equation}
\label{cal-L-local}
 \mathcal{L}_0 (\lambda) = \sum _{m=1}^N \left( \frac{\vec{\sigma}_{0} \cdot \vec{S}_{m}}{\lambda - \alpha _m}  +    \frac{ \vec{\sigma}_{0} \cdot  \left( K _m^{-1}(\lambda) \vec{S}_{m} K _m(\lambda)\right)}{\lambda + \alpha _m}   \right).
\end{equation}
The Gaudin Hamiltonians with the boundary terms are obtained from the residues of the generating function \eqref{open-tau} at  poles $\lambda = \pm\alpha_m$ :
\begin{equation} 
\operatorname{Res}_{\lambda = \alpha_m} \tau (\lambda) \ =\  4 \, H_m \quad
%- 4\alpha_m( 1 + \phi \psi )\,\vec{S}_{m} \cdot \vec{S}_{m}\\
\text{and} \quad 
\operatorname{Res}_{\lambda = -\alpha_m} \tau (\lambda) \ =\ 4\, \widetilde H_m
%+ 4\alpha_m( 1 + \phi \psi )\,\vec{S}_{m} \cdot \vec{S}_{m}
\end{equation}
where
\begin{equation}
\label{open-Ham-a}
H_m = \sum _{n \neq m}^N \frac{\vec{S}_{m} \cdot \vec{S}_{n}}{\alpha _m - \alpha _n} + 
\sum _{n = 1}^N  \frac{ \left( K _m (\alpha_m) \vec{S}_{m} K _m^{-1}  (\alpha_m) \right)  \cdot  \vec{S}_{n} + \vec{S}_{n} \cdot  \left( K _m (\alpha_m) \vec{S}_{m} K _m^{-1}  (\alpha_m) \right) }{2(\alpha_m + \alpha _n)} ,
\end{equation}
and 
\begin{equation}
\label{open-Ham-b}
\widetilde{H}_m = \sum _{n \neq m}^N \frac{\vec{S}_{m} \cdot \vec{S}_{n}}{\alpha _m - \alpha _n} +  
\sum _{n = 1}^N  \frac{ \left( K _m (-\alpha_m) \vec{S}_{n} K _m^{-1} (-\alpha_m) \right)  \cdot  \vec{S}_{n} + \vec{S}_{n} \cdot  \left( K _m (-\alpha_m) \vec{S}_{m} K _m^{-1} (-\alpha_m) \right) }{2(\alpha_m + \alpha _n)} .
\end{equation}

Since the element $\Delta \left[\mathcal{T}(\lambda)\right]$ can be written in form \eqref{Del-calT}
it is evident that the vector $\Omega _+$ \eqref{Omega+} is its eigenvector
\begin{equation}
\label{Del-calTOm}
\Delta \left[\mathcal{T}(\lambda)\right] \Omega _+ =  2 \lambda \, \alpha (\lambda + \eta / 2) \, \widehat{\delta} (\lambda - \eta / 2) \Omega _+ .
\end{equation}
Moreover, it follows from \eqref{t-on-Om+} and \eqref{Del-calTOm} that $\Omega _+$ \eqref{Omega+}  is an eigenvector of the difference
\begin{equation}
\label{t-DonOm}
\left( 2 \lambda t(\lambda) - \Delta \left[\mathcal{T}(\lambda)\right] \right) \Omega _+ = 2 \lambda 
\left( \Lambda _0 (\lambda) - \alpha (\lambda + \eta / 2) \, \widehat{\delta} (\lambda - \eta / 2) \right) \Omega _+ .
\end{equation}
We can expand the eigenvalue on the right hand side of the equation above in powers of $\eta$
\begin{align}
\label{exp-chi0}
&2\lambda \left(  \kappa _1 (\lambda) \alpha (\lambda) + \kappa _2 (\lambda) \widehat{\delta} (\lambda) - \alpha (\lambda + \eta / 2) \, \widehat{\delta} (\lambda - \eta / 2) \right) 
= 2\lambda  \left(\xi^2 - \lambda ^2 \nu ^2 \right)  \notag \\
&+ \eta \left(\xi^2 - 3 \lambda ^2 \nu ^2 \right) + \eta ^2  \lambda \left( \left( \xi ^2 - \lambda ^2  \nu ^2 \right)  \chi _0 (\lambda) - \frac{\nu ^2}{2} \right)+ \mathcal{O}(\eta ^3) .  
\end{align}
Substituting the expansion above into the right hand side of \eqref{t-DonOm} and using \eqref{final-exp-t-D-open} to expand the left hand side, it follows that the vector $\Omega _+$ \eqref{Omega+} is an eigenvector of the generating function of the Gaudin Hamiltonians
\begin{equation}
\label{egnv-chi0}
\tau (\lambda) \Omega _+ = \chi _0 (\lambda)  \Omega _+,
\end{equation}
with
\begin{equation}
\label{chi0}
\begin{split}
\chi _0 (\lambda) &= \frac{4 \lambda}{\xi ^2 - \lambda ^2  \nu ^2} \sum_{m=1}^N \left( \frac{s_m}{\lambda - \alpha _m} + \frac{s_m}{\lambda + \alpha _m} \right) \\
&+2 \sum _{m,n = 1}^N  \left( \frac{ s _m s _n + s_m \delta _{mn} }{(\lambda - \alpha _m) (\lambda - \alpha _n)}  + \frac{2\left( s _m s _n + s_m \delta _{mn}  \right)}{(\lambda - \alpha _m) (\lambda + \alpha _n)} + \frac{s _m s _n + s_m \delta _{mn}}{(\lambda + \alpha _m) (\lambda + \alpha _n)} \right) .
\end{split}
\end{equation}
As expected, the eigenfunction $\chi _0 (\lambda)$ also depends on the boundary parameters $\xi, \nu$. In general we can obtain the spectrum $\chi _M (\lambda , \mu _1, \dots , \mu _M)$ of the generating function $\tau (\lambda)$ of the Gaudin Hamiltonians  through the expansion
\begin{align}
\label{exp-Lambda}
&2\lambda \left(  \Lambda _M (\lambda , \mu _1, \dots , \mu _M) - \alpha (\lambda + \eta / 2) \, \widehat{\delta} (\lambda - \eta / 2) \right) = 2\lambda  \left(\xi^2 - \lambda ^2 \nu ^2 \right)  \notag \\
&+ \eta \left(\xi^2 - 3 \lambda ^2 \nu ^2 \right) + \eta ^2  \lambda \left( \left( \xi ^2 - \lambda ^2  \nu ^2 \right) \chi _M (\lambda , \mu _1, \dots , \mu _M) - \frac{\nu ^2}{2} \right) + \mathcal{O}(\eta ^3) ,  
\end{align}
or explicitly 
\begin{equation}
\label{chi-M}
\begin{split}
&\chi _M (\lambda , \mu _1, \dots , \mu _M) = \frac{-4\lambda ^2 \nu ^4}{(\xi ^2 - \lambda ^2 \nu ^2) ^2} + 2 \sum _{j,k = 1}^M  \left( \frac{ 1 - \delta _{jk} }{(\lambda - \mu _j) (\lambda - \mu _k)}  + 
\frac{2\left( 1 - \delta _{jk}   \right)}{(\lambda -\mu _j) (\lambda +  \mu _k)} 
+ \frac{1 - \delta _{jk} }{(\lambda + \mu _j) (\lambda +  \mu _k)} \right) \\[2ex]
&+ 2 \sum _{m,n = 1}^N  \left( \frac{ s _m s _n + s_m \delta _{mn} }{(\lambda - \alpha _m) (\lambda - \alpha _n)} + \frac{2\left( s _m s _n + s_m \delta _{mn}  \right)}{(\lambda - \alpha _m) (\lambda + \alpha _n)} + \frac{s _m s _n + s_m \delta _{mn}}{(\lambda + \alpha _m) (\lambda + \alpha _n)} \right) \\[2ex]
&- 4 \left( \sum_{j=1}^M \left( \frac{1}{\lambda - \mu _j} + \frac{1}{\lambda + \mu _j}\right) - \frac{\lambda \nu ^2}{\xi ^2 - \lambda ^2 \nu ^2} \right) \left( \sum_{m=1}^N \left( \frac{s_m}{\lambda - \alpha _m} + \frac{s_m}{\lambda + \alpha _m} \right) + \frac{\lambda \nu ^2}{\xi ^2 - \lambda ^2 \nu ^2} \right) .
\end{split}
\end{equation}

As our next important step toward obtaining the formulas of the algebraic Bethe ansatz for the corresponding Gaudin model we observe that the first term in the expansion of the function $F_M(\mu_1; \mu_2, \ldots, \mu_M)$ in powers of $\eta$ is  
\begin{equation}
\label{Fandf}
F_M(\mu_1; \mu_2, \ldots, \mu_M) =  \eta f_M (\mu_1; \mu_2, \ldots, \mu_M)  + \mathcal{O}(\eta ^2) ,
\end{equation}
where
\begin{equation}
\label{f-M}
\begin{split}
f_M (\mu_1; \mu_2, \ldots, \mu_M)  &= \frac{2 \mu _1\nu^2}{\xi + \mu _1 \nu} - 2 (\xi - \mu _1 \nu) \sum _{j =2}^M \left( \frac{1}{\mu _1 - \mu _j} + \frac{1}{\mu _1 +  \mu _j} \right) \\
&+ 2 (\xi - \mu _1 \nu) \sum _{m=1}^N \left( \frac{s_m}{\mu _1 - \alpha _m} + \frac{s_m}{\mu _1 + \alpha _m} \right) .
\end{split}
\end{equation}

We have used the formulas \eqref{Psi1} and \eqref{b1} as well as \eqref{CalB} and \eqref{deltahat} in order to expand the Bethe vector $\Psi _1(\mu)$ of the Heisenberg spin chain in powers of $\eta$ and obtained the Bethe vector $\varphi_1 (\mu)$ of the Gaudin model
\begin{equation}
\label{Psi1-exp}
\Psi _1(\mu) = \eta \varphi_1 (\mu)  + \mathcal{O}(\eta ^2) , 
\end{equation}
where
\begin{equation}
\label{phi1}
\varphi_1 (\mu) = \sum _{m=1}^N \left(  \frac{ \xi  + \alpha _m \nu }{\mu  - \alpha _m} + \frac{\xi  + \alpha _m \nu} {\mu  + \alpha _m } \right) \left( \frac{\psi \, s_m}{\nu}  + S^-_m \right)
\Omega _+ .
\end{equation}
As our final step we observe that using \eqref{Del-calT} and \eqref{t-on-Psi1} we have the off shell action of the difference of the transfer matrix of the XXX chain and the central element, the so-called Sklyanin determinant, on the Bethe vector $\Psi _1 (\mu)$
\begin{equation}
\label{t-DonPsi1}
\begin{split}
\left( 2 \lambda t(\lambda) - \Delta \left[\mathcal{T}(\lambda)\right] \right) \Psi _1 (\mu) &= 2 \lambda 
\left( \Lambda _1 (\lambda, \mu) - \alpha (\lambda + \eta / 2) \, \widehat{\delta} (\lambda - \eta / 2) \right) \Psi _1 (\mu) \\
&+ (2 \lambda) \frac{2 \eta (\lambda + \eta)(\xi + \mu \nu)}{(\lambda - \mu) (\lambda + \mu + \eta)} F_1 (\mu) \Psi _1 (\lambda) .
\end{split}
\end{equation}
Finally, the off shell action of the generating function the Gaudin Hamiltonians on the vector $\varphi_1 (\mu)$ can be obtained from the equation above by using the expansion \eqref{final-exp-t-D-open} and \eqref{Psi1-exp} on the left hand side as well as the expansion \eqref{exp-Lambda}, \eqref{Fandf} and \eqref{Psi1-exp} 
on the right hand side 
\begin{equation}
\label{tau-on-phi1}
\begin{split}
\tau (\lambda) \varphi_1 (\mu) &= \chi _1 (\lambda, \mu)  \varphi_1 (\mu) +  \frac{4 \lambda (\xi + \mu \nu)}{(\xi ^2 - \lambda ^2  \nu ^2)(\lambda ^2 -\mu ^2)} f_1(\mu) \varphi_1 (\lambda) .
\end{split}
\end{equation}
Therefore $\varphi_1 (\mu)$ \eqref{phi1} is the Bethe vector of the corresponding Gaudin model, i.e.  the eigenvector of the generating function the Gaudin Hamiltonians once the unwanted term is canceled by imposing the corresponding Bethe equation
\begin{equation}
\label{GM-BE1}
 f_1(\mu) = \frac{2 \mu \, \nu^2}{\xi + \mu \nu} + 2 (\xi - \mu \nu) \sum _{m=1}^N \left( \frac{s_m}{\mu - \alpha _m} + \frac{s_m}{\mu + \alpha _m} \right) = 0.
\end{equation}

To obtain the action of the generating function $\tau (\lambda)$ on the Bethe vector $\varphi_2 (\mu_1,\mu_2)$ of the Gaudin model we follow analogous steps to the ones we have done when studding the action of $\tau (\lambda)$ on $\varphi_1 (\mu)$. The first term in the expansion of the Bethe vector $\Psi _2 ( \mu_1 ,  \mu_2)$ \eqref{Psi2} in powers of $\eta$ yields the corresponding Bethe vector of the Gaudin model
\begin{equation}
\label{Psi2-exp}
\Psi _2(\mu _1 , \mu_2 ) = \eta ^2 \varphi_2 (\mu_1,\mu_2)  + \mathcal{O}(\eta ^3) , 
\end{equation}
where
\begin{equation}
\label{phi2}
\begin{split}
&\varphi_2 (\mu_1,\mu_2) = \sum _{m,n=1}^N   \left(  \frac{ \xi  + \alpha _m \nu }{\mu _1  - \alpha _m} + \frac{\xi  + \alpha _m \nu} {\mu _1 + \alpha _m } \right)  \left(  \frac{ \xi  + \alpha _n \nu }{\mu _2  - \alpha _n} + \frac{\xi  + \alpha _n \nu} {\mu _2 + \alpha _n } \right) \times \\[1ex]
&\times 
\left( \left( \frac{\psi \, s_m}{\nu}  + S^-_m \right) \left( \frac{\psi \, s_n}{\nu}  + S^-_n \right) -  \frac{\psi}{\nu} \delta _{mn} \left( \frac{\psi \, s_n}{2\nu}  + S^-_n \right) \right) \Omega _+ .
\end{split}
\end{equation}
As in the previous case \eqref{t-DonPsi1}, it is of interest to study the action of the difference of the transfer matrix $t(\lambda)$ and the so-called Sklyanin determinant $\Delta \left[\mathcal{T}(\lambda)\right]$ on the Bethe vector $\Psi _2(\mu _1 , \mu_2 )$ using \eqref{Del-calT} and \eqref{t-on-Psi2} 
\begin{equation}
\label{t-DonPsi2}
\begin{split}
\left( 2 \lambda t(\lambda) - \Delta \left[\mathcal{T}(\lambda)\right] \right) \Psi _2(\mu _1 , \mu_2 ) &= 2\lambda \left(  \Lambda _2 (\lambda , \mu _1 , \mu _2) - \alpha (\lambda + \eta / 2) \, \widehat{\delta} (\lambda - \eta / 2) \right)\Psi _2(\mu _1 , \mu_2 ) \\[1ex]
&+ (2\lambda) \frac{2 \eta (\lambda + \eta ) (\xi + \mu _1 \nu )}{(\lambda - \mu _1) (\lambda + \mu _1 + \eta )} F_2 (\mu _1; \mu _2) \Psi _2 ( \lambda ,  \mu_2) \\[1ex]
&+ (2\lambda) \frac{2 \eta (\lambda + \eta ) (\xi + \mu _2 \nu )}{(\lambda - \mu _2) (\lambda + \mu _2 + \eta )} F_2(\mu _2; \mu _1) \Psi _2 ( \lambda ,  \mu_1) .
\end{split}
\end{equation}
The off shell action of the generating function of the Gaudin Hamiltonians on the Bethe vector $\varphi_2 (\mu_1,\mu_2)$ is obtained from the equation above using the expansions \eqref{final-exp-t-D-open} and \eqref{Psi2-exp} on the left hand side and \eqref{exp-Lambda}, \eqref{Psi2-exp} and 
\eqref{Fandf} on the right hand side. Then, by comparing the terms of the fourth power in $\eta$ on both sides of \eqref{t-DonPsi2} we derive  
\begin{equation}
\label{tau-on-phi2}
\begin{split}
\tau (\lambda) \varphi_2 (\mu_1, \mu_2) &= \chi _2 (\lambda , \mu _1, \mu _2) \varphi_2 (\mu_1,\mu_2) +  \frac{4 \lambda (\xi + \mu _1 \nu)}{(\xi ^2 - \lambda ^2  \nu ^2)(\lambda ^2 -\mu _1^2)} f_2 (\mu_1; \mu_2) \varphi_2 (\lambda, \mu_2) \\[1ex]
&+  \frac{4 \lambda (\xi + \mu _2 \nu)}{(\xi ^2 - \lambda ^2  \nu ^2)(\lambda ^2 -\mu _2^2)} f_2 (\mu_2; \mu_1) \varphi_2 (\lambda, \mu_1) .
\end{split}
\end{equation}
The two unwanted terms on the right hand side of the equation above are annihilated by the following Bethe equations 
\begin{align}
\label{GM-BE2-1}
f_2 (\mu_1; \mu_2)  &= \frac{2 \mu _1\nu^2}{\xi + \mu _1 \nu} - 2 (\xi - \mu _1 \nu) \left( \frac{1}{\mu _1 - \mu _2} + \frac{1}{\mu _1 +  \mu _2} \right) 
+ 2 (\xi - \mu _1 \nu) \sum _{m=1}^N \left( \frac{s_m}{\mu _1 - \alpha _m} + \frac{s_m}{\mu _1 + \alpha _m} \right) = 0 , \\[1ex]
\label{GM-BE2-2}
f_2 (\mu_2; \mu_1)  &= \frac{2 \mu _2\nu^2}{\xi + \mu _2 \nu} - 2 (\xi - \mu _2 \nu) \left( \frac{1}{\mu _2 - \mu _1} + \frac{1}{\mu _2 +  \mu _1} \right)
+ 2 (\xi - \mu _2 \nu) \sum _{m=1}^N \left( \frac{s_m}{\mu _2 - \alpha _m} + \frac{s_m}{\mu _2 + \alpha _m} \right) = 0 .
\end{align}
The off shell action of the generating function $\tau (\lambda)$ on the Bethe vector $\varphi_2 (\mu_1,\mu_2)$ of the Gaudin model is strikingly simple \eqref{tau-on-phi2}. Actually, it is as simple as it can be since \eqref{tau-on-phi2} practically coincide with the corresponding formula in the case when the boundary matrix $K(\lambda)$ is diagonal \cite{Hikami95}. 

In general, we have that the first term in the expansion of the Bethe vector \break $\Psi _M ( \mu_1 ,  \mu_2 ,  \dots ,  \mu_M )$ \eqref{PsiM}, for arbitrary positive integer $M$, in powers of $\eta$ is
\begin{equation}
\label{PsiM-exp}
\Psi _M ( \mu_1 ,  \mu_2 ,  \dots ,  \mu_M ) = \eta ^M \varphi_M ( \mu_1 ,  \mu_2 ,  \dots ,  \mu_M )  + \mathcal{O}(\eta ^{M+1}) , 
\end{equation}
where 
\begin{equation}
\label{phiM}
\varphi_M ( \mu_1 ,  \mu_2 ,  \dots ,  \mu_M ) = F (\mu _1) F (\mu _2) \cdots F (\mu _M)  \Omega _+
\end{equation}
and the operator $F (\mu)$ is given by
\begin{equation}
\label{operatorF}
F (\mu ) = \sum _{m=1}^N  \left( \frac{\xi + \mu \nu}{\mu - \alpha _m} +  \frac{\xi - \mu \nu}{\mu + \alpha _m}\right)  \left( \frac{\psi}{\nu} S^3_m + S^-_m - \frac{\psi ^2}{4\nu ^2}  S^+_m \right) .
\end{equation}
The Bethe vector of the Gaudin model $\varphi_M ( \mu_1 ,  \mu_2 ,  \dots ,  \mu_M )$ is a symmetric function of its arguments, since a straightforward calculation shows that the operator $F (\mu )$  
commutes at different values of the spectral parameter,
\begin{equation}
\label{F-commute}
[F (\lambda) , F (\mu)] = 0.
\end{equation}
The action of the generating function $\tau (\lambda)$ on the Bethe vector $\varphi_M ( \mu_1 ,  \mu_2 ,  \dots ,  \mu_M )$ is derived analogously to the previous two cases when $M = 1$ \eqref{tau-on-phi1} and $M = 2$ \eqref{tau-on-phi2}. In the present case we use the expansions 
\eqref{exp-Lambda}, \eqref{Fandf} and \eqref{PsiM-exp} to obtain 
\begin{equation}
\label{tau-on-phiM}
\begin{split}
\tau (\lambda) \varphi_M ( \mu_1 ,  \mu_2 ,  \dots ,  \mu_M ) &= \chi _M(\lambda , \{\mu _i \} _{i=1}^M) \varphi_M ( \mu_1 ,  \mu_2 ,  \dots ,  \mu_M ) \\
&+ \sum _{i=1}^M \frac{4 \lambda (\xi + \mu _i \nu)}{(\xi ^2 - \lambda ^2  \nu ^2)(\lambda ^2 -\mu _i^2)}
f_M (\mu_i; \{\mu _j \} _{j \neq i} ) \varphi_M ( \lambda ,  \{\mu _j \} _{j \neq i} ) ,
\end{split}
\end{equation}
where $\chi _M(\lambda , \{\mu _i \} _{i=1}^M)$ is given in \eqref{chi-M} and the unwanted terms on the right hand side of the equation above are canceled by the following Bethe equations 
\begin{equation}
\label{GM-BEM-i}
\begin{split}
f_M (\mu_i; \{\mu _j \} _{j \neq i} )  &= \frac{2 \mu _i\nu^2}{\xi + \mu _i \nu} - 2 (\xi - \mu _i \nu) \sum _{\substack{ j =1\\ j \neq i}}^M \left( \frac{1}{\mu _i - \mu _j} + \frac{1}{\mu _i +  \mu _j} \right) \\
&+ 2 (\xi - \mu _i \nu) \sum _{m=1}^N \left( \frac{s_m}{\mu _i - \alpha _m} + \frac{s_m}{\mu _i + \alpha _m} \right) = 0 ,
\end{split}
\end{equation}
for $i = 1 , 2 , \dots M$. As expected, the above action of the generating function $\tau (\lambda)$ is
strikingly simple and this simplicity is due to our definition of the Bethe vector $\varphi_M ( \mu_1 ,  \mu_2 ,  \dots ,  \mu_M )$ \eqref{phiM}. These results will be studied further in the framework of an alternative approach to the implementation of the algebraic Bethe ansatz for the Gaudin model, with triangular triangular K-matrix \eqref{K-min-GM}, based on the classical reflection equation and corresponding linear bracket and will be reported in \cite{CAMRS}.

\section{Conclusions}
We have implemented fully the off shell algebraic Bethe ansatz for the XXX Heisenberg spin chain in the case when the boundary parameters satisfy an extra condition guaranteeing that both boundary matrices  can be brought to the upper-triangular form by a single similarity matrix which does not depend on the spectral parameter. As it turned out the identity satisfied by the Lax operator enables a convenient  realization for the Sklyanin monodromy matrix.  This realization led to the action of the entries of the Sklyanin monodromy matrix on the vector $\Omega _+$ and consequently to the observation that $\Omega _+$ is an eigenvector of the transfer matrix of the chain.

We have proceeded then to the essential step of the algebraic Bethe ansatz, to the definition of the Bethe vectors $\Psi _M ( \mu_1 ,  \mu_2 ,  \dots ,  \mu_M )$. Our objective was to make the off shell action of the transform matrix $t(\lambda)$ on them as simple as possible. Before defining the general Bethe vector $\Psi _M ( \mu_1 ,  \mu_2 ,  \dots ,  \mu_M )$, for an arbitrary positive integer $M$, we gave a step by step presentation of the first four Bethe vectors, including the formulae for the action of $t(\lambda)$, the corresponding eigenvalues and Bethe equations. In this way we have exposed the striking property of these vectors to make the off shell action of the transform matrix as simple as possible. Consequently, the elaborated definition of $\Psi _M ( \mu_1 ,  \mu_2 ,  \dots ,  \mu_M )$, for arbitrary positive integer $M$, appeared naturally as a generalization of the first four Bethe vectors. 
As expected, the action of $t(\lambda)$ on the Bethe vector $\Psi _M ( \mu_1 ,  \mu_2 ,  \dots ,  \mu_M )$ is again very simple. Actually, the action of the transfer matrix is as simple as it could possible be since it almost coincides with the corresponding action in the case when the two boundary matrices are diagonal \cite{Sklyanin88, Hikami95}. 

We explored further these results by obtaining the off shell action of the generating function of the Gaudin Hamiltonians on the corresponding Bethe vectors  by means of the so-called quasi-classical limit. To study the open Gaudin model we had to impose the condition so that the parameters of the reflection matrices on the left and on the right end of the chain are the same. This is not the case in  the study of the open spin chain, but is essential for the Gaudin model. The generating function of the Gaudin Hamiltonians with boundary terms is derived analogously to the periodic case \cite{CAMRS}. Based on this result we showed how the quasi-classical limit yields the off shell action of the generating function of the Gaudin Hamiltonians on the Bethe vectors $\varphi_M ( \mu_1 ,  \mu_2 ,  \dots ,  \mu_M )$ as well as the spectrum and the Bethe equations. The off shell action of the generating function $\tau (\lambda)$ on the Bethe vectors $\varphi_M ( \mu_1 ,  \mu_2 ,  \dots ,  \mu_M )$ is strikingly simple. As in the case of the spin chain, it is as simple as it can be since it  practically coincide with the corresponding formula in the case when the boundary matrix is diagonal \cite{Hikami95}. This simplicity of the action of $\tau (\lambda)$ is due to our definition of the Bethe vectors $\varphi_M ( \mu_1 ,  \mu_2 ,  \dots ,  \mu_M )$. 

An important open problem is to calculate the off shell scalar product of the Bethe vectors we have defined above both for the XXX Heisenberg spin chain and the Gaudin model. These results could lead to the correlations functions for both systems. In the case of Gaudin model it would be of interest to establish a relation between Bethe vectors and solutions of the corresponding Knizhnik-Zamolodchikov, along the lines it was done in the case  when the boundary matrix is diagonal \cite{Hikami95}. 

\bigskip

\noindent
\textbf{Acknowledgments}

\noindent
We acknowledge useful discussions with Eric Ragoucy and Zolt\'an Nagy. I. S. was\break supported in part by the Serbian Ministry of Science and Technological Development under grant number ON 171031. N. M. is thankful to Professor Victor Kac and the staff of the Mathematics Department at MIT for their  warm hospitality. N. M. was supported in part by the FCT sabbatical fellowship SFRH/BSAB/1366/2013.

\clearpage
\newpage

\appendix
\section{Basic definitions}
We consider the spin operators 
$S^{\alpha}$ 
with $\alpha = +, - , 3$, acting in some (spin $s$) representation space $\mathbb{C}^{2s+1}$ with the commutation relations
\begin{equation}
\label{crspin1}
[S^3, S^{\pm}] = \pm S^{\pm}, \quad [S^+,S^-] = 2 S^3 , 
\end{equation}
and Casimir operator
$$c_2 = (S^3) ^2 + \frac{1}{2} (S^+S^-+S^-S^+) = (S^3) ^2 + S^3 + S^-S^+=\vec{S}\cdot\vec{S}.$$
In the particular case of spin $\frac12$ representation, one recovers the Pauli matrices
$$
S^{\alpha} = \frac{1}{2} \sigma ^{\alpha} = \frac{1}{2} \left(\begin{array}{cc}
\delta_{\alpha3} & 2\delta_{\alpha+}  \\
2\delta_{\alpha-} & - \delta _{\alpha 3} \end{array}\right).
$$ 

We consider a spin chain with N sites with spin $s$ representations, i.e. a local $\mathbb{C}^{2s+1}$ space at each site and the operators 
\begin{equation}
S_m^{\alpha} = \mathbbm{1} \otimes \cdots \otimes \underbrace{S^{\alpha}} _m \otimes \cdots \otimes \mathbbm{1},
\end{equation}
with $\alpha = +,-, 3$ and $m= 1, 2 ,\dots , N$.

\section{Commutation relations} 
The equation \eqref{exchangeRE} yields the exchange relations between the operators $\mathcal{A} (\lambda)$, $\mathcal{B} (\lambda)$, $\mathcal{C} (\lambda)$  and $\widehat{\mathcal{D}} (\lambda)$.
The relevant relations are
\begin{align}
\label{comm-relBB+CC}
\mathcal{B} (\lambda) \mathcal{B} (\mu) &= \mathcal{B} (\mu) \mathcal{B} (\lambda) , \qquad  
\mathcal{C} (\lambda) \mathcal{C} (\mu) = \mathcal{C} (\mu) \mathcal{C} (\lambda) ,  \\[1ex]
\label{comm-relAB}
\mathcal{A} (\lambda) \mathcal{B} (\mu) &= 
\frac{(\lambda + \mu)(\lambda - \mu - \eta)}{(\lambda - \mu)(\lambda + \mu + \eta)} \mathcal{B} (\mu) \mathcal{A} (\lambda) + \frac{2\eta\mu}{(\lambda - \mu)(2 \mu + \eta)} \mathcal{B} (\lambda) \mathcal{A} (\mu) \notag \\
&- \frac{\eta}{\lambda +\mu + \eta} \mathcal{B} (\lambda) \widehat{\mathcal{D}} (\mu) , \\[1ex]
\label{comm-rel-hDB}
\widehat{\mathcal{D}} (\lambda) \mathcal{B} (\mu) &= \frac{(\lambda - \mu + \eta)(\lambda + \mu + 2 \eta)}{(\lambda - \mu) (\lambda + \mu + \eta)} \mathcal{B} (\mu) \widehat{\mathcal{D}} (\lambda) - \frac{2\eta (\lambda + \eta)} {(\lambda - \mu)(2 \lambda + \eta)} \mathcal{B} (\lambda) \widehat{\mathcal{D}} (\mu) \notag \\
&+ \frac{4 \eta \mu (\lambda + \eta)}{(2\lambda + \eta)(2\mu + \eta)(\lambda +\mu + \eta)} \mathcal{B} (\lambda) \mathcal{A} (\mu) , \\[1ex]
\label{comm-relCB}
\left[ \mathcal{C} (\lambda)  , \mathcal{B} (\mu) \right] &= \frac{2 \eta \lambda (\lambda - \mu + \eta)}{(\lambda - \mu) (\lambda + \mu + \eta) (2 \lambda + \eta)} \mathcal{A} (\mu) \mathcal{A} (\lambda) - \frac{2 \eta ^2 \lambda}{(\lambda - \mu) (2 \lambda + \eta) (2 \mu + \eta)} \mathcal{A} (\lambda) \mathcal{A} (\mu)  \notag \\
&+\frac{\eta (\lambda + \mu) }{(\lambda-\mu)(\lambda + \mu + \eta)} \mathcal{A} (\mu) \widehat{\mathcal{D}} (\lambda)  - \frac{2 \eta \lambda}{(\lambda - \mu) (2 \lambda + \eta)} \mathcal{A} (\lambda) \widehat{\mathcal{D}} (\mu) \notag \\
&-\frac{\eta ^2}{(\lambda + \mu + \eta) (2 \mu + \eta)}\widehat{\mathcal{D}} (\lambda) \mathcal{A} (\mu) - \frac{\eta}{\lambda + \mu + \eta} \widehat{\mathcal{D}} (\lambda) \widehat{\mathcal{D}} (\mu) .
\end{align}
For completeness we include the following commutation relations 
\begin{align}
\label{AA}
\left [ \mathcal{A} (\lambda) , \mathcal{A} (\mu) \right] &= \frac{\eta}{\lambda + \mu + \eta}  \left( \mathcal{B} (\mu) \mathcal{C} (\lambda) - \mathcal{B} (\lambda) \mathcal{C} (\mu) \right) \\
\label{AD}
\left [ \mathcal{A} (\lambda) , \widehat{\mathcal{D}} (\mu)  \right] &= \frac{2\eta (\mu + \eta)}{(\lambda - \mu)(2\mu + \eta)} \left( \mathcal{B} (\lambda) \mathcal{C} (\mu) - \mathcal{B} (\mu) \mathcal{C} (\lambda) \right) \\
\label{DD}
\left [ \widehat{\mathcal{D}} (\lambda) , \widehat{\mathcal{D}} (\mu)  \right] &= \frac{4\eta (\lambda + \eta) (\mu + \eta)}{(2\lambda + \eta)(2\mu + \eta) (\lambda + \mu + \eta)} \left( \mathcal{B} (\lambda) \mathcal{C} (\mu) - \mathcal{B} (\mu) \mathcal{C} (\lambda) \right)
\end{align}
From the relations above it follows that 
\begin{equation}
\label{ABBOmega}
\begin{split}
\mathcal{A} (\lambda) \mathcal{B} (\mu_1) \mathcal{B} (\mu_2) \Omega _+
&= \prod _{i = 1}^2
\frac{(\lambda + \mu_i)(\lambda - \mu_i - \eta)}{(\lambda - \mu _i)(\lambda + \mu_i + \eta)} 
\alpha (\lambda) \mathcal{B} (\mu_1) \mathcal{B} (\mu_2)  \Omega _+ \\
&+ \sum _{i=1} ^2 \frac{2 \eta \mu_i}{(2 \mu _i + \eta)(\lambda - \mu_i)} \frac{(\mu_i + \mu_{3-i})(\mu_i  - \mu_{3-i} - \eta)}{(\mu_i - \mu _{3-i})(\mu_i + \mu_{3-i} + \eta)} \alpha (\mu_i) \mathcal{B} (\lambda) \mathcal{B} (\mu_{3-i})  \Omega _+ \\
&- \sum _{i=1} ^2 \frac{\eta}{\lambda + \mu_i + \eta} \frac{(\mu_i - \mu_{3-i} + \eta)(\mu_i + \mu_{3-i} + 2\eta)}{(\mu_i - \mu _{3-i})(\mu_i + \mu_{3-i} + \eta)}  \widehat{ \delta} (\mu_i) \mathcal{B} (\lambda) \mathcal{B} (\mu_{3-i}) \Omega _+  .
\end{split}
\end{equation}
Analogously, 
\begin{equation}
\label{DBBOmega}
\begin{split}
\widehat{\mathcal{D}} (\lambda) \mathcal{B} (\mu_1) \mathcal{B} (\mu_2) \Omega _+
&= \prod _{i = 1}^2
\frac{(\lambda - \mu_i + \eta)(\lambda + \mu_i + 2 \eta )}{(\lambda - \mu_i )(\lambda + \mu _i + \eta)}  
\widehat{ \delta} (\lambda) \mathcal{B} (\mu_1) \mathcal{B} (\mu_2)  \Omega _+ \\
&-  \sum _{i=1} ^2 \frac{2 \eta (\lambda + \eta )}{(2 \lambda + \eta)(\lambda - \mu_i)} \frac{(\mu_i  - \mu_{3-i} + \eta)(\mu_i + \mu_{3-i} + 2 \eta)}{(\mu_i - \mu _{3-i})(\mu_1 + \mu_{3-i} + \eta)} \widehat{ \delta} (\mu_i) \mathcal{B} (\lambda) \mathcal{B} (\mu_{3-i}) \Omega _+ \\
&+ \sum _{i=1} ^2 \frac{4\eta \mu_i (\lambda + \eta)}{(2 \lambda + \eta)(2 \mu _i + \eta)(\lambda + \mu_i + \eta)} \times \\
& \times 
\frac{(\mu_i + \mu_{3-i} )(\mu_i - \mu_{3-i} - \eta)}{(\mu_i - \mu _{3-i})(\mu_i + \mu_{3-i} + \eta)} \alpha (\mu_i) \mathcal{B} (\lambda) \mathcal{B} (\mu_{3-i}) \Omega _+ . 
\end{split}
\end{equation}
Finally, 
\begin{align}
\label{CBBOmega}
&\mathcal{C} (\lambda) \mathcal{B} (\mu_1) \mathcal{B} (\mu_2)\Omega _+ =  \sum _{i=1}^2
\left( 
\frac{4 \mu_i \lambda \eta}{(2 \lambda + \eta)(2 \mu_i + \eta)(\lambda + \mu _i + \eta)} \times 
\right. \notag \\
&\times \frac{(\lambda + \mu_{3-i})(\lambda - \mu_{3-i} - \eta)}{(\lambda - \mu_{3-i})(\lambda + \mu_{3-i} + \eta)}\frac{(\mu_i + \mu_{3-i})(\mu_i - \mu_{3-i} - \eta)}{(\mu_i - \mu_2)(\mu_i + \mu_{3-i} + \eta)}
\alpha (\lambda) \alpha (\mu_i) 
- \frac{2 \lambda \eta}{(\lambda -  \mu _i)(2 \lambda + \eta)} \times 
\notag \\
&\times \frac{(\lambda + \mu_2)(\lambda - \mu_2 - \eta)}{(\lambda - \mu_2)(\lambda + \mu_2 + \eta)} \frac{(\mu_i - \mu_2 + \eta)(\mu_i + \mu_2 + 2 \eta)}{(\mu_i - \mu_2)(\mu_i + \mu_2 + \eta)} 
\alpha (\lambda) \widehat{\delta} (\mu_i) 
+ \frac{2 \mu_i \eta}{(\lambda -  \mu _i)(2 \mu_i + \eta)} \times
\notag \\
&\times \frac{(\lambda - \mu_2 + \eta)(\lambda + \mu_2 + 2 \eta)}{(\lambda - \mu_2)(\lambda + \mu_2 + \eta)} 
\frac{(\mu_i + \mu_2)(\mu_i - \mu_2 - \eta)}{(\mu_i - \mu_2)(\mu_i + \mu_2 + \eta)}
 \alpha (\mu_i) \widehat{\delta} (\lambda) 
%
%& 
- \frac{\eta}{\lambda + \mu _i + \eta} \times 
\notag \\
& \left. \times \frac{(\lambda - \mu_2 + \eta)(\lambda + \mu_2 + 2 \eta)}{(\lambda - \mu_2)(\lambda + \mu_2 + \eta)} \frac{(\mu_i - \mu_2 + \eta)(\mu_i + \mu_2 + 2 \eta)}{(\mu_i - \mu_2)(\mu_i + \mu_2 + \eta)} \widehat{\delta} (\lambda) \widehat{\delta} (\mu_1) \right)
\mathcal{B} (\mu_{3-i}) \Omega _+ \notag \\
&+ \left( \frac{8 \eta ^2  \mu_1 \mu_2 \, (\mu_1 + \mu_2)(\lambda (\lambda + \eta) - \mu_1 \mu_2)}{(\lambda - \mu_1)(\lambda - \mu_2) (2 \mu_1 + \eta) (2 \mu_2 + \eta) (\lambda + \mu_1 + \eta) (\lambda + \mu_2 + \eta) (\mu_1 + \mu_2 + \eta)} 
\alpha (\mu _1) \alpha (\mu _2)  \right. \notag \\
&- \frac{4 \eta ^2 \mu_1 \, (\mu_2 - \mu_1 + \eta) ( \lambda (\lambda + \eta) + \mu_1 (\mu_2 + \eta) )}{(\lambda - \mu _1) (\lambda - \mu _2) (2 \mu _1 + \eta) (\mu _2 - \mu _1)(\lambda + \mu_1 + \eta) (\lambda + \mu_2 + \eta)} 
\alpha (\mu_1)  \widehat{\delta} (\mu_2)  \notag \\
&- \frac{4 \eta ^2 \mu_2 \, (\mu_1 - \mu_2 + \eta) ( \lambda (\lambda + \eta) + \mu_2 (\mu_1 + \eta) ) }{(\lambda - \mu_1)(\lambda - \mu_2) (2 \mu_2 + \eta) (\mu_1 - \mu_2) (\lambda + \mu_1 + \eta) (\lambda + \mu_2 + \eta) } 
\alpha (\mu _2) \widehat{\delta} (\mu_1)  \notag \\
&\left. - \frac{2 \eta ^2  (\mu _1 + \mu_2 + 2 \eta) (\eta ^2 - \lambda ^2 + \mu _1 \mu_ 2 + \eta (\mu _1 + \mu_2 - \lambda) ) }{(\lambda - \mu_1)(\lambda - \mu_2) (\lambda + \mu_1 + \eta) (\lambda + \mu_2 + \eta) (\mu _1 + \mu_2 + \eta) }
\widehat{\delta} (\mu_1)  \widehat{\delta} (\mu_2) \right) \mathcal{B} (\lambda) \Omega _+ 
\end{align}
The relations \eqref{ABBOmega}, \eqref{DBBOmega} and \eqref{CBBOmega} are readily generalized 
\cite{Eric13}.

\clearpage
\newpage

\end{document}